\documentclass[twocolumn]{aastex631}

\usepackage{amsmath}

\begin{document}

\title{Retrieval of Thermally-Resolved Water Vapor Distributions in Disks Observed with JWST-MIRI}

\correspondingauthor{Carlos E. Romero-Mirza}
\email{carlos.romero\_mirza@cfa.harvard.edu}

\author[0000-0001-7152-9794]{Carlos E. Romero-Mirza}
\affiliation{Center for Astrophysics $\vert$ Harvard \& Smithsonian, Cambridge, MA 02138, USA}

\author[0000-0003-4335-0900]{Andrea Banzatti}
\affiliation{Department of Physics, Texas State University, 749 N Comanche Street, San Marcos, TX 78666, USA}

\author[0000-0001-8798-1347]{Karin I. Öberg}
\affiliation{Center for Astrophysics $\vert$ Harvard \& Smithsonian, Cambridge, MA 02138, USA}

\author[0000-0001-7552-1562]{Klaus M. Pontoppidan}
\affiliation{Jet Propulsion Laboratory, California Institute of Technology, 4800 Oak Grove Dr. Pasadena, CA, 91109, USA}

\author[0000-0003-3682-6632]{Colette Salyk}
\affiliation{Vassar College, 124 Raymond Avenue, Poughkeepsie, NY 12604, USA}

\author[0000-0002-5758-150X]{Joan Najita}
\affiliation{NSFs NOIRLab, 950 N. Cherry Avenue, Tucson, AZ 85719, USA}

\author[0000-0003-0787-1610]{Geoffrey A. Blake}
\affiliation{Division of Geological \& Planetary Sciences, California Institute of Technology, Pasadena, CA 91125, USA}

\author[0000-0002-3291-6887]{Sebastiaan Krijt}
\affiliation{School of Physics and Astronomy, University of Exeter, Stocker Road, Exeter, EX4 4QL, UK}

\author[0000-0003-2631-5265]{Nicole Arulanantham}
\affiliation{Space Telescope Science Institute, 3700 San Martin Drive, Baltimore, MD 21218, USA}

\author[0000-0001-8764-1780]{Paola Pinilla}
\affiliation{Mullard Space Science Laboratory, University College London, Holmbury St Mary, Dorking, Surrey, RH5 6NT, UK}

\author[0000-0002-7607-719X]{Feng Long}
\affiliation{Lunar and Planetary Laboratory, University of Arizona, Tucson, AZ 85721, USA}
\altaffiliation{NASA Hubble Fellowship Program Sagan Fellow}

\author[0000-0003-4853-5736]{Giovanni Rosotti}
\affiliation{Dipartimento di Fisica, Università degli Studi di Milano, Via Celoria 16, 20133 Milano, Italy}

\author[0000-0003-2253-2270]{Sean M. Andrews}
\affiliation{Center for Astrophysics $\vert$ Harvard \& Smithsonian, Cambridge, MA 02138, USA}

\author[0000-0003-1526-7587]{David J. Wilner}
\affiliation{Center for Astrophysics $\vert$ Harvard \& Smithsonian, Cambridge, MA 02138, USA}

\author[0000-0002-0150-0125]{Jenny Calahan}
\affiliation{Center for Astrophysics $\vert$ Harvard \& Smithsonian, Cambridge, MA 02138, USA}

\author{The JDISCS Collaboration}

\begin{abstract}

The mid-infrared water vapor emission spectrum provides a novel way to characterize the delivery of icy pebbles towards the innermost ($<5$ au) regions of planet-forming disks. Recently, JWST MIRI-MRS showed that compact disks exhibit an excess of low-energy water vapor emission relative to extended multi-gapped disks, suggesting that icy pebble drift is more efficient in the former. We carry out detailed emission line modeling to retrieve the excitation conditions of rotational water vapor emission in a sample of four compact and three extended disks within the JDISC Survey. We present two-temperature H$_2$O slab model retrievals and, for the first time, constrain the spatial distribution of water vapor by fitting parametric radial temperature and column density profiles. Such models statistically outperform the two-temperature slab fits. We find a correlation between the observable hot water vapor mass and stellar mass accretion rate, as well as an anti-correlation between cold water vapor mass and sub-mm dust disk radius, confirming previously reported water line flux trends. We find that the mid-IR spectrum traces H$_2$O with temperatures down to 180-300 K, but the coldest 150-170 K gas remains undetected. Furthermore the H$_2$O temperature profiles are generally steeper and cooler than the expected `super-heated' dust temperature in passive irradiated disks. The column density profiles are used to estimate icy pebble mass fluxes, which suggest that compact and extended disks may produce markedly distinct inner-disk exoplanet populations if local feeding mechanisms dominate their assembly.

\end{abstract}

\section{Introduction} \label{sec:intro}

Ever since the first detection of water vapor in a protoplanetary disk by \citet{Carr_2004} \citep[see also][]{Carr_2008, Salyk_2008}, there has been considerable interest in using its infrared emission spectrum as a tracer of the thermochemical and dynamical properties of the innermost $(< 5 \text{ au} )$ disk regions. One particularly interesting trend that emerged from the \textit{Spitzer InfraRed Spectrograph (Spitzer-IRS)} observations is an anti-correlation between sub-mm dust disk masses and mid-IR H$_2$O line luminosities relative to those of organics \citep{Carr_2011, Najita_2013}. This was interpreted as evidence that more massive disks are more efficient at forming icy planetesimals, which would end up locking up most of the O-rich ice. Using resolved continuum imaging with ALMA, a follow-up study by \citet{Banzatti_2020} in turn indicated that the strongest anti-correlation occurs between mid-IR H$_2$O line fluxes and the sub-mm pebble disk radius---and this correlation persists even after accounting for the accretion luminosity of the host. Hydro-dynamical processes have provided a physical explanation for such a trend: if, unlike multi-gapped extended disks, compact disks are created by efficient pebble drift \citep[e.g.][]{Pinilla_2012,appelgren20, Toci2021, vanderMarel_2021}, brighter H$_2$O line emission could indeed be observed as due to the increased rate of ice sublimation near the snowline \citep{Cyr_1998, Ciesla_2006, kalyaan21, Kalyaan2023}.

A more direct observational test to determine whether compact disks exhibit efficient pebble drift can be obtained by measuring the excess emission of moderate-energy (with upper-level energies $E_{up} \lesssim 3000$ K) rotational water lines. This is because such H$_2$O transitions are expected to be in near local thermo-dynamic equilibrium (LTE), and can be dominated by emission from gas temperatures between 200-400 K \citep{Salyk_2011, Du_2017, Banzatti_2020, Banzatti_2023a} consistent with the expected thermal conditions in elevated layers near the snowline, while warm ($>$ 400 K) gas-phase water formation would dominate the emission from higher energy levels \citep{Glassgold_2009, Meijerink_2009, Blevins_2016, Bosman_2022}. Measuring the relative excitation of lower- vs higher-energy water lines was not possible earlier with $\textit{Spitzer-IRS}$ spectra, since most mid-IR H$_2$O lines were blended together at its low spectral resolving power ($R \sim 600$). Instead, with JWST MIRI-MRS (henceforth MIRI-MRS) \citep{Wells_2015, Argyriou_2023} it is now possible to carry out spectroscopic studies of water vapor with unprecedented detail thanks to its sensitivity, wavelength coverage, and resolving power ($2000-4000$) \citep[e.g.][]{Grant_2023, Gasman_2023, Pontoppidan_2024, Xie_2023}. Indeed, MIRI-MRS can now be used to search for direct evidence as to whether compact disks exhibit brighter low-energy H$_2$O emission lines compared to extended disks \citep{Banzatti_2023b}.

The aim of this work is to go beyond comparing the H$_2$O spectral line patterns between compact and extended disks \citet{Banzatti_2023b}, and carry out detailed emission line modeling of the entire rotational spectra to retrieve the excitation conditions of water vapor in a sample of compact and extended disks with a number of dust gaps. We present a novel technique to fit H$_2$O emission spectra using not only single- \citep[e.g.][]{Grant_2023} or two-temperature \citep{Pontoppidan_2024, Romero_Mirza_2024} slabs, but also using physically-motivated parametric models to characterize the radial temperature and column density structure of water vapor in each disk. It is known that single slab models cannot reproduce the entire rotational H$_2$O emission spectrum of disks due to excitation temperature $(T_{ex})$ gradients \citep{Blevins_2016, Liu_2019, Banzatti_2023a, Gasman_2023}, but so far radial gradients have rarely been included in mid-IR molecular excitation retrievals \citep{Zhang_2013, Kaeufer_2024}. By modeling temperature and density gradients, we can start to use H$_2$O emission as a proxy to study the inner disk physical structure---from otherwise spatially- and kinematically-unresolved observations. Extracting radial distributions, furthermore, makes it possible to provide estimates for the inner-disk icy pebble mass fluxes, which in turn can give clues as to which kind of planets can form via the pebble accretion mechanism across localized radii in the disk \citep{Ida_2016, Lambrechts_2019}.  

This paper is organized as follows: in Section \ref{sec:obs}, we present the MIRI-MRS data used in this paper. Section \ref{sec:mod_and_res} presents the slab modeling technique and Bayesian retrieval details, as well as their results. We discuss our findings in Section \ref{sec:dis} and, finally, present our conclusions in Section \ref{sec:conc}.


\section{Observations} \label{sec:obs}

We analyze new and published MIRI-MRS observations of seven protoplanetary disks around T Tauri stars. Four systems have compact pebble disks (with resolved sub-mm dust disk radii $R_{dust}<30$ au) while three are significantly more extended ($R_{dust}>100$ au). Our sample includes the four H$_2$O-rich disks first presented by \citet{Banzatti_2023b}: GK Tau, CI Tau, IQ Tau, and HP Tau, which provided the first confirmed cases of cold water vapor excess in compact disks. In addition, we analyze the compact disk of FZ Tau, known to harbor an abundant reservoir of hot and cold water vapor \citep{Pontoppidan_2024}, and the large disk of AS 209, which exhibits bright dust continuum emission and a mid-IR spectrum dominated by hot H$_2$O \citet{Romero_Mirza_2024}. Finally, we present new MIRI-MRS observations of the compact disk around the classical T Tauri  star (cTTs) GQ Lup. 

The disk of GQ Lup was observed as part of JWST Cycle 1 program GO-1640 (P.I. Andrea Banzatti) on August 13, 2023, with deep integrations of $1700-1800$ s per MRS sub-band, with the specific goal of characterizing the degree of inner-disk water vapor enrichment. For observational details regarding the rest of our sample we direct the reader to \citet{Banzatti_2023b}, \citet{Pontoppidan_2024}, and \citet{Romero_Mirza_2024}. The MIRI-MRS data of all disks in our sample were processed using the JWST Disk Infrared Spectral Chemistry Survey (JDISCS) reduction pipeline version 7.1, which includes an empirical de-fringing process using asteroid calibrators. This is fully described by \citet{Pontoppidan_2024}. Table \ref{tab:sample} summarizes the properties of the disks in our sample.

\begin{figure*}[ht!]
    \centering
    \includegraphics[width=0.98\textwidth]{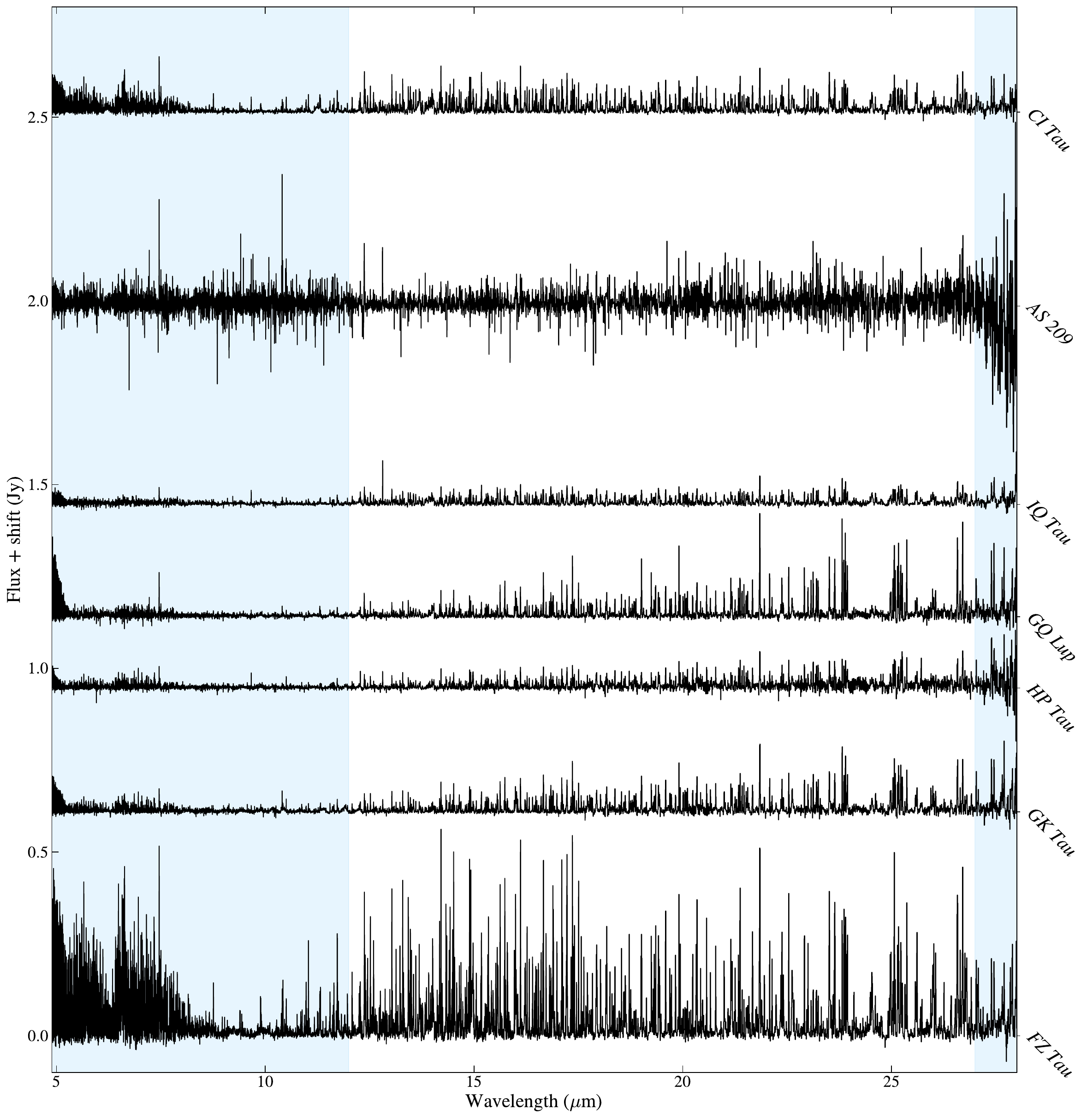}
    \caption{Continuum-subtracted MIRI-MRS spectra of each of the disks in the sample, vertically shifted for clarity. The blue shaded regions indicate the wavelength regions excluded from our analysis due to ro-vibrational emission ($< 12$ $\mu$m) or larger noise ($> 27$ $\mu$m).}
    \label{fig:contsub_spectra}
\end{figure*}

Figure \ref{fig:contsub_spectra} presents the continuum-subtracted spectra for all the disks included in this study, and the wavelength region considered in our analysis. The dust continuum is subtracted by fitting a Gaussian Process model to the data, as described in \citet{Romero_Mirza_2024}. We also experiment with using a Savitzky–Golay filter \citep[e.g.][]{Pontoppidan_2024} to subtract the continuum emission, and find negligible differences. A plot showing full spectra and empirical continuum models for each disk is presented in Appendix \ref{apx:full}.

In this paper, we focus on the analysis of water vapor rotational emission from each of these disks. The identification and retrieval of column densities and excitation conditions of organics and other species in these and other sources within the JDISCS survey will be presented in an upcoming overview paper (Arulanantham et al., in prep.). 

\begin{deluxetable*}{lcccccccl}
\label{tab:sample}
\tabletypesize{\small}
\tablecolumns{9}
\tablewidth{0pt}
\tablecaption{Properties of the disk sample}
\tablehead{
\colhead{Source} & \colhead{Distance} & \colhead{$i$} & \colhead{$R_{dust}$} & \colhead{$L_{\star}$} & \colhead{$M_{\star}$} & \colhead{$\log_{10} \dot{M}$} & \colhead{Innermost gap} & \colhead{JWST Program} \\
\colhead{} & \colhead{(pc)} & \colhead{(deg)} & \colhead{(au)} & \colhead{$(L_{\odot})$} & \colhead{$(M_{\odot})$} & \colhead{$(M_{\odot} yr^{-1})$} & \colhead{(au)} & \colhead{}
}
\startdata  
FZ Tau & 129 & 25 & 12 & 0.97 & 0.56 & -6.5 & -- & GO1 1549 (P.I. Klaus Pontoppidan) \\
GK Tau & 129 & 39 & 13 & 1.48 & 0.58 & -8.7 & -- & GO1 1640 (P.I. Andrea Banzatti) \\
HP Tau & 177 & 18 & 22 & 1.89 & 0.84 & -8.2 & $2^{a}$ & GO1 1640 \\
GQ Lup & 154 & 61 & 24 & 1.44 & 0.70 & -7.9 & 8$^{b}$ & GO1 1640 \\
IQ Tau & 131 & 62 & 110 & 0.86 & 0.42 & -8.5 & 41 & GO1 1640 \\
AS 209 & 121 & 35 & 140 & 1.41 & 1.20 & -7.3 & 9 & GO1 2025 (P.I. Karin Öberg) \\
CI Tau & 160 & 50 & 190 & 1.65 & 0.65 & -7.6 & 14$^{c}$ & GO1 1640 \\
\enddata
\tablecomments{Distances obtained from GAIA DR3 \citet{Gaia_2023}. Disk inclinations $(i)$ obtained from \citet{MacGregor_2017, Long_2019, Pontoppidan_2024}. Sub-mm dust disk radii $(R_{dust})$ obtained from \citet{Huang_2018b, Long_2019, Long_2020, Wu_2017}, and correspond to the radius enclosing $95\%$ of the emission. Stellar luminosities $(L_{\star})$ and masses $(M_{\star})$ obtained from \citet{Andrews_2018, Herczeg_2014, alcala_2017, McClure_2019}. Mass accretion rates obtained from \citet{Salyk_2013, Herczeg_2014, alcala_2017, McClure_2019}. Innermost dust gap locations from \citet{Huang_2018b, Long_2018, Long_2020}. $a$ Putative cavity based on the near infrared index \citep[see discussion in][]{Banzatti_2023b}. $b$ The inner gap found in GQ Lup is very shallow and may not produce an effective trap against pebble drift \citep{Kalyaan2023}. $c$ An additional inner gap at 5 au has been proposed by \citet{jennings22_taurus} using super-resolution techniques.}
\end{deluxetable*}

\section{Modeling and Results} \label{sec:mod_and_res}

We aim to retrieve the physical properties of water vapor emitting from the innermost regions of the seven disks in our sample. The mid-IR emission spectrum of H$_2$O is generated using the LTE slab modeling Python package \texttt{iris} \citep{Romero_Mirza_InfraRed_Isothermal_2023} with spectroscopic data obtained from the HITRAN database \citep{Gordon_2022}. The slab models consist of opacity-weighted intensities that account for line blending and line-center saturation effects, which is critical to obtain accurate column density measurements. The opacity-weighted H$_2$O line intensities are modeled assuming a line Gaussian width given by the sum in quadrature of a 1 km/s turbulent component and a thermal broadening component $\sqrt{k_BT_{ex}/\mu}$, where $\mu$ is the molecular weight of H$_2$O. Typical thermal widths are of 0.4-0.8 km/s for the best-fit temperatures reported below. We only consider in our modeling the $12.0 - 27.0$ $\mu$m wavelength range, since emission at shorter wavelengths is dominated by high-energy ro-vibrational H$_2$O lines likely emitting from hot and diffuse disk regions where water vapor is not thermalized by collisions with H$_2$ \citep{Carr_2004, Meijerink_2009, van_Dishoeck_2014, Banzatti_2023a}. \texttt{iris} has been benchmarked against the publicly available package \texttt{spectools\_ir} \citep{slabspec}.

The slab models are calculated assuming a different resolving power within the wavelength coverage of each MIRI-MRS sub-band. First, we generate the line emission models at a sufficiently high resolution ($R\approx 10^5$) to sample the line profiles. Then, the models are convolved with a Gaussian kernel of width $\Delta_{\lambda} = \lambda_m/R$, where $\lambda_m$ is the median wavelength of each sub-band, and binned onto a wavelength grid with uniform spacing $\Delta_{\lambda}/2$. To calculate the log-likelihood of the models, the MIRI-MRS data and their uncertainty in each sub-band is binned onto the same uniform wavelength grid while ensuring flux conservation using the Python package \texttt{SpectRes} \citep{Carnall_spectres}. Appendix \ref{apx:noise} describes how the flux uncertainty in each sub-band is estimated. Table \ref{tab:resolving_power} lists the resolving power and wavelength sampling assumed for each instrument sub-band. We emphasize that the full $12.0 - 27.0$ $\mu$m region is modeled simultaneously.

\begin{deluxetable}{ccc}
\label{tab:resolving_power}
\tabletypesize{\small}
\tablecolumns{3}
\tablewidth{0pt}
\tablecaption{Slab model details}
\tablehead{
\colhead{Region ($\mu$m)} & \colhead{Resolving Power} & \colhead{$\lambda$ Sampling ($\mu$m)}}
\startdata  
$12.0 - 13.5$ & 2950 & $ 2.2\times 10^{-3}$ \\
$13.5 - 15.5$ & 2725 & $ 2.7\times 10^{-3}$ \\
$15.5 - 18.0$ & 2780 & $ 3.0\times 10^{-3}$ \\
$18.0 - 21.0$ & 2220 & $ 4.4\times 10^{-3}$ \\
$21.0 - 24.5$ & 2235 & $ 5.1\times 10^{-3}$ \\
$24.5 - 27.0$ & 1960 & $ 6.6\times 10^{-3}$ \\
\enddata
\tablecomments{The resolving power and wavelength sampling used to model the water vapor emission in each wavelength region of interest. The values of $R$ are estimated from the empirical fits reported by \citet{Pontoppidan_2024}.}
\end{deluxetable}

The modeling and retrieval details are presented below. In brief, we first model the H$_2$O emission in each disk as the sum of two slabs to represent two temperature components: a first approximation to a radial temperature gradient as done e.g. in \citet{Pontoppidan_2024}. Next, we fit power law excitation temperature and column density profiles to each disk. Finally, based on the results of the two previous retrievals, we fit a more flexible model with a tapered power law column density profile. All models are fit using the Bayesian nested sampling Python package \texttt{dynesty} \citep{Speagle_2020}. We apply multiple ellipsoid decomposition and random walk sampling, and use the change in remaining evidence as a measure of convergence.

\subsection{Two-Temperature Water Vapor Models}

Observations with \textit{Spitzer-IRS}, \textit{Herschel-PACS}, and ground observatories first provided indications that the IR spectrum of water vapor from disks cannot be typically reproduced by a slab model with a single excitation temperature, but rather multiple temperature components are necessary \citep[for detailed overviews see][]{van_Dishoeck_2014, Banzatti_2023a}. This has recently been confirmed with slab model fits to MIRI-MRS disk spectra \citep{Banzatti_2023b, Gasman_2023, Xie_2023, Pontoppidan_2024}, finding that models with two temperature components can reproduce the rotational H$_2$O line emission reasonably well. We first model the H$_2$O spectra for the disks in our sample following this simpler approximation.

Specifically, the spectra are modeled as the sum of two plane-parallel slabs, each with an emitting area ($A$), water column density $(N)$, and excitation temperature $(T_{ex})$ \citep{Romero_Mirza_2024}. We use the following Uniform (in $\log_{10}$ space) prior distributions for $A$ and $N$:

\begin{equation}
\begin{split}
\log_{10} N &= \mathcal{U} (15, 20) \text{ cm}^{-2}\\
\log_{10} A &= \mathcal{U} (-3, 3) \text{ au}^{2},
\label{eq:uniform}
\end{split}
\end{equation}
and the following Normal priors for the excitation temperatures:
\begin{equation}
\begin{split}
\log_{10} T_{cold} &= \mathcal{N} (2.6, 0.2) \text{ K}\\
\log_{10} T_{hot} &= \mathcal{N} (2.9, 0.1) \text{ K}\\
\label{eq:normal}
\end{split}
\end{equation}
for the cold and hot water vapor components, respectively. These correspond to Normal priors centered around 400 and 800 K, with standard deviation of $\sim 200$ K. Note that the use of Normal priors does not affect the best-fit parameters in any meaningful way, and they are simply used to consistently guide one slab towards the low- and another towards the high-temperature solution. The priors are informed by previous best-fit water vapor parameters reported by \citet{Banzatti_2023b}, \citet{Pontoppidan_2024}, and \citet{Romero_Mirza_2024} for six of the disks analyzed in this work. 

\begin{figure}[ht!]
    \centering
    \includegraphics[width=0.45\textwidth]{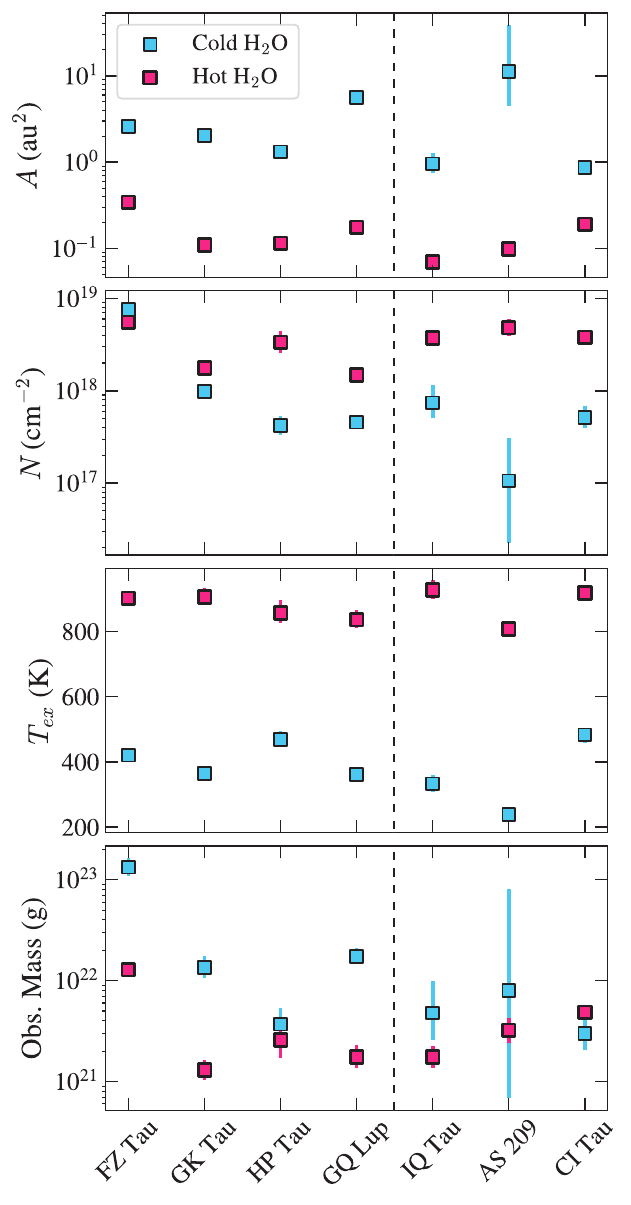}
    \caption{Best-fit two-temperature H$_2$O model parameters for each disk in our sample. Error bars correspond to the 0.16 and 0.84 percentiles. Observable masses are calculated as the product of column density, emitting area, and molecular weight. The disks are listed in ascending $R_{dust}$ order from left to right, and the dashed lines separate compact and extended disks . Note that sometimes the error bars are smaller than the marker.}
    \label{fig:2comp_params}
\end{figure}

Figure \ref{fig:2comp_params} presents the best-fit emitting areas, column densities, and excitation temperatures for each disk. Here and in the rest of this paper, the `best-fit' values are chosen to be the median of the posterior distribution obtained for each parameter. Table \ref{tab:2c_results} lists the exact values and corresponding uncertainties for each parameter. We find that while the emitting areas of the cold and hot water vapor reservoirs are different from each other, they are each similar across the sample. The cold H$_2$O components tend to have emitting areas of $1- 10$ au$^2$ (or equivalent radii $R_{eq} = 0.6 - 1.8$ au assuming circular areas), whereas the hot components have areas of order $0.1$ au$^2$ ($R_{eq} = 0.2$ au). The hot H$_2$O column densities are also fairly similar for compact and extended disks, with typical values of $ 10^{18}-10^{19}$ cm$^{-2}$. On average, however, we find that compact disks exhibit higher cold H$_2$O column densities than extended disks, by a factor of a few. With respect to the excitation temperatures, we retrieve comparable $T_{ex}$ values across the sample of  $800-1000$ K for hot H$_2$O and $\sim 400$ K for cold H$_2$O, with some variation: HP Tau and CI Tau have warmer cold H$_2$O components ($\sim 500$ K), while AS 209 exhibits a very cold H$_2$O component ($\sim 250$ K). 

Note that the poor quality of the AS 209 spectrum affects the robustness of this particular fit. In agreement with \citet{Romero_Mirza_2024}, we find that there are multiple high-energy H$_2$O lines with sufficient SNR to accurately constrain the hot water vapor component. Yet, while detected, the lower-energy H$_2$O lines are affected by noise artifacts beyond 20 $\mu$m. This results in a significantly higher relative uncertainty for the retrieved emitting area and column density of cold water vapor in AS 209 compared to the rest of the sample (see Figure \ref{fig:2comp_params}). The excitation temperature, however, appears to be well constrained by the line ratios. As further discussed in Section \ref{sec:tempas209}, while the peculiar properties of cold H$_2$O in AS 209 may arise due to the lower SNR, they could also indicate the presence of inner disk substructure, for which there is kinematic evidence \citep{Banzatti_2023a}.

\begin{deluxetable*}{lcccccl}
\label{tab:2c_results}
\tabletypesize{\small}
\tablecolumns{9}
\tablewidth{0pt}
\tablecaption{Two-component H$_2$O fit results}
\tablehead{
\colhead{Source} & \colhead{$N_{cold}/10^{18}$} & \colhead{$N_{hot}/10^{18}$} & \colhead{$T_{cold}$} & \colhead{$T_{hot}$} & \colhead{$A_{cold}$} & \colhead{$A_{hot}$} \\
\colhead{} & \colhead{(cm$^{-2}$)} & \colhead{(cm$^{-2}$)} & \colhead{(K)} & \colhead{(K)} & \colhead{(au$^{2}$)} & \colhead{(au$^2$)} 
}
\startdata  
FZ Tau & $7.64^{+1.39}_{-1.12}$ & $5.57^{+0.60}_{-0.52}$ & $420^{+10}_{-9}$ & $903^{+15}_{-23}$ & $2.60^{+0.11}_{-0.11}$ & $0.34^{+0.03}_{-0.02}$ \\ 
GK Tau & $0.98^{+0.18}_{-0.15}$ & $1.77^{+0.23}_{-0.22}$ & $365^{+12}_{-11}$ & $905^{+27}_{-25}$ & $2.04^{+0.20}_{-0.16}$ & $0.11^{+0.01}_{-0.01}$ \\
HP Tau & $0.42^{+0.11}_{-0.09}$ & $3.34^{+1.11}_{-0.77}$ & $468^{+28}_{-23}$ & $858^{+40}_{-32}$ & $1.32^{+0.17}_{-0.14}$ & $0.11^{+0.02}_{-0.02}$ \\
GQ Lup & $0.46^{+0.06}_{-0.05}$ & $1.48^{+0.25}_{-0.19}$ & $361^{+7}_{-8}$ & $836^{+28}_{-26}$ & $5.60^{+0.44}_{-0.34}$ & $0.18^{+0.02}_{-0.02}$ \\
IQ Tau & $0.74^{+0.41}_{-0.23}$ & $3.75^{+0.64}_{-0.56}$ & $333^{+27}_{-24}$ & $926^{+31}_{-27}$ & $0.96^{+0.33}_{-0.21}$ & $0.07^{+0.01}_{-0.01}$ \\
AS 209 & $0.11^{+0.21}_{-0.08}$ & $4.85^{+1.12}_{-0.90}$ & $238^{+23}_{-18}$ & $807^{+25}_{-22}$ & $11.44^{+28.74}_{-6.91}$ & $0.10^{+0.01}_{-0.01}$ \\
CI Tau & $0.52^{+0.17}_{-0.12}$ & $3.79^{+0.40}_{-0.35}$ & $483^{+21}_{-25}$ & $917^{+17}_{-16}$ & $0.86^{+0.11}_{-0.10}$ & $0.19^{+0.01}_{-0.01}$ \\
\enddata
\tablecomments{The reported values for each parameter consist of the median and 0.16-0.84 posterior distribution percentiles.}
\end{deluxetable*}

Given the variation in excitation temperatures among hot and cold components in the sample, we caution against over-interpreting apparent correlations (or lack thereof) between H$_2$O column densities and/or areas and disk sizes at this stage. That is, some of the water vapor classified as `cold' in some disks would contribute to the `hot' component in others. While one could in principle alleviate this by fitting the data with fixed-temperature components (e.g. one with $T_{ex} = 400$ and one with $T_{ex} = 800$ K), we find that doing so produces significantly larger residuals. This, combined with the fact that we do not observe strong correlations between $T_{ex}$, $A$, and $N$, seems to indicate that the source-to-source variation in retrieved temperatures occurs due to distinct radial temperature structures between disks.

\begin{figure*}[ht!]
    \centering
    \includegraphics[width=0.98\textwidth]{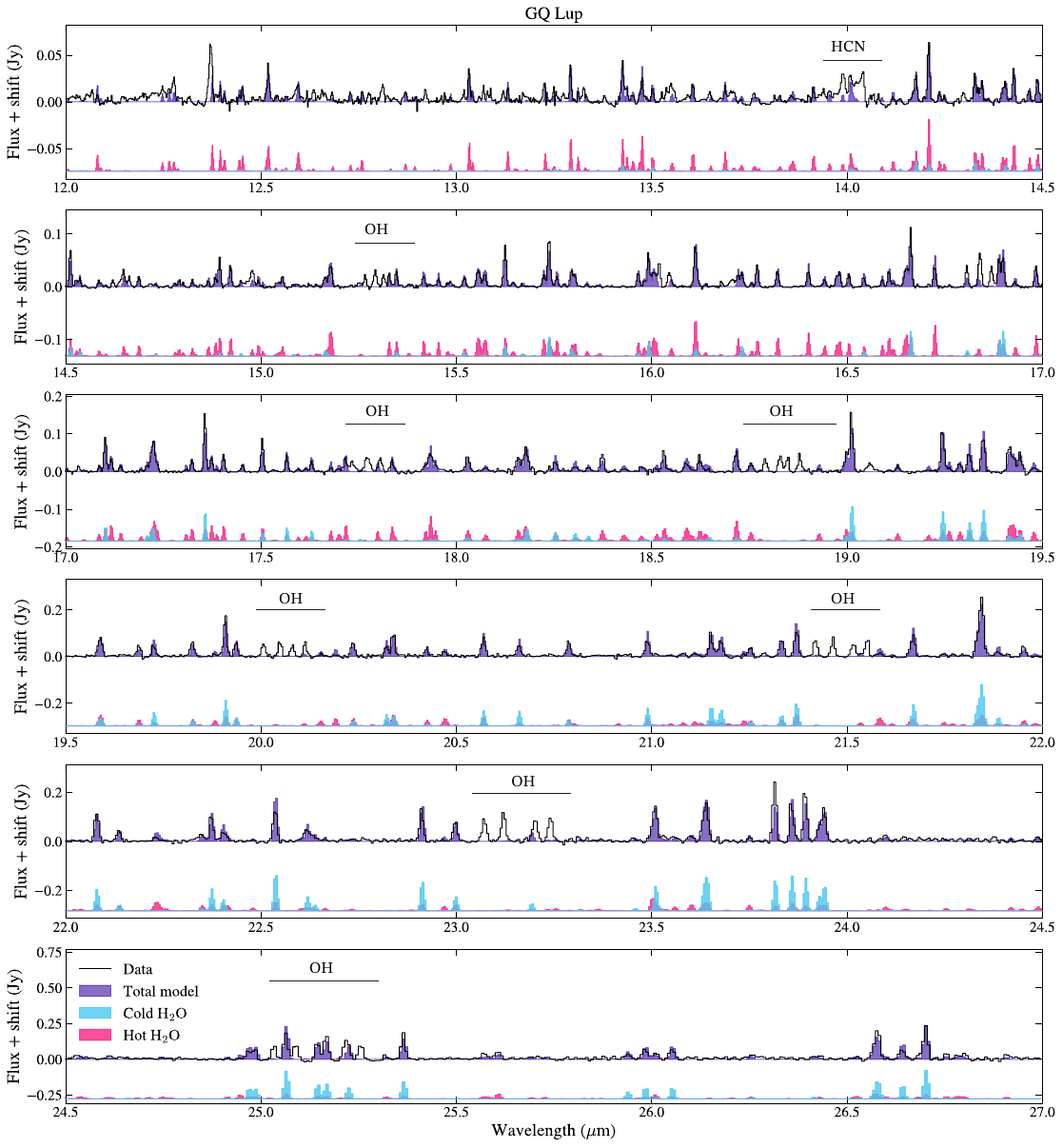}
    \caption{Best-fit two-temperature H$_2$O model to the compact disk of GQ Lup. The data is shown in black, the total model in violet, and the individual contributions of the cold and hot H$_2$O components are shown in blue and magenta, respectively, shifted vertically for clarity. Regions where the emission is blended with OH or organics are labeled.}
    \label{fig:gqlup_2comp}
\end{figure*}

\begin{figure*}[ht!]
    \centering
    \includegraphics[width=0.98\textwidth]{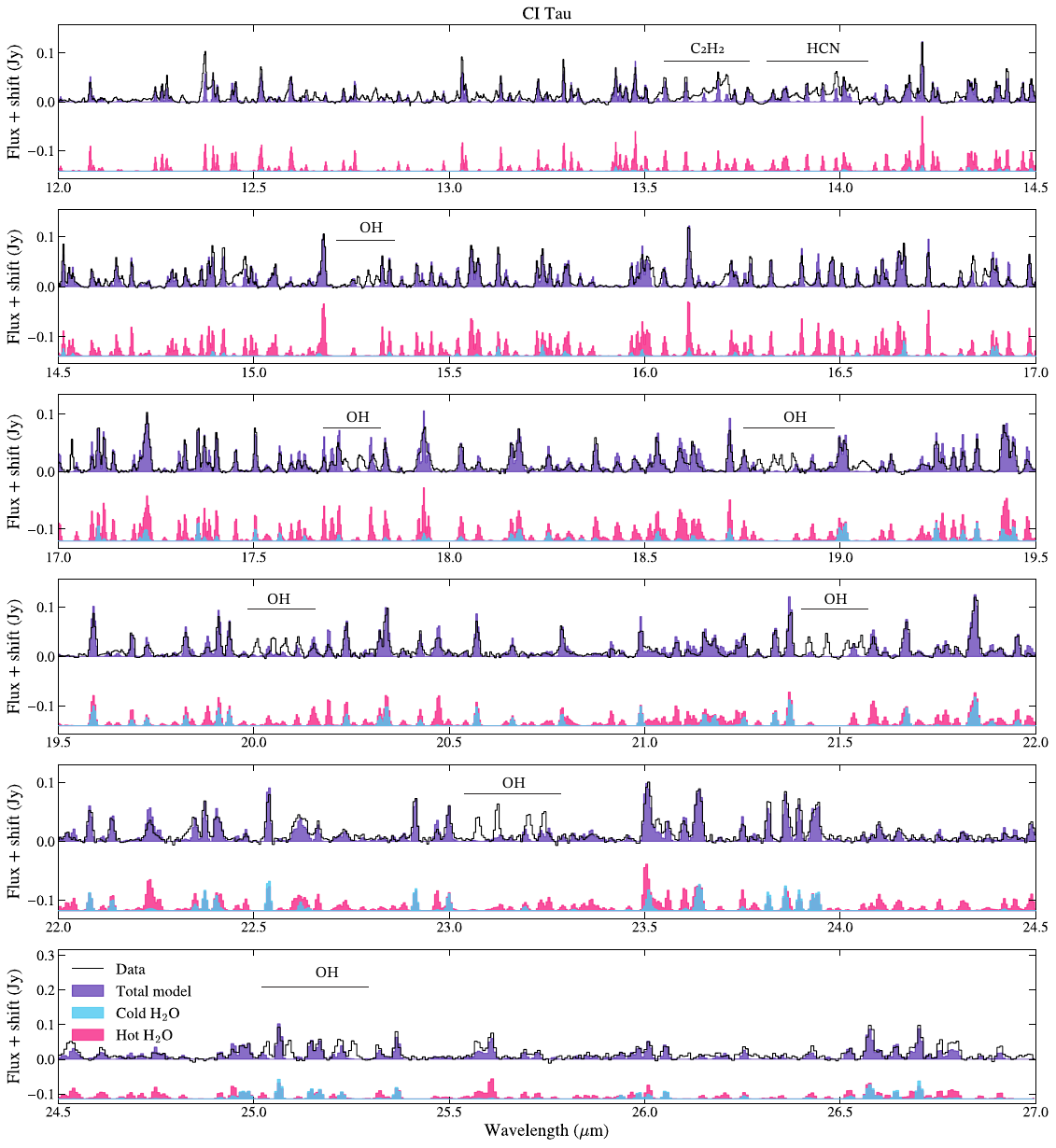}
    \caption{Same as Figure \ref{fig:gqlup_2comp}, for the extended disk of CI Tau. Note that unlike GQ Lup, the spectrum of CI Tau is dominated by the hot H$_2$O component even at longer wavelengths.}
    \label{fig:citau_2comp}
\end{figure*}

Figure \ref{fig:gqlup_2comp} shows the best-fit two-temperature H$_2$O model for the compact disk of GQ Lup. The contributions of the hot and cold components are shown individually, and the spectrum has been broken up into multiple panels for clarity. We emphasize that including a cold H$_2$O component is critical to reproduce the low-energy water lines beyond 20 $\mu$m. Note that most of the bright lines not reproduced by the model (e.g. near 16.8, 18.8, 20.1, 21.5, 23.2, and 25.3 $\mu$m) are OH emission lines. By contrast, Figure \ref{fig:citau_2comp} presents the best-fit two-temperature H$_2$O model to the extended disk of CI Tau. Note how in this case, as reported in \citet{Banzatti_2023b}, the spectrum is dominated by hot H$_2$O even at long wavelengths, with little excess from colder H$_2$O emission. The two-temperature models fit to the rest of the sample are presented in Appendix \ref{apx:2t}.   

\subsection{Temperature and Column Density Gradients}
It is likely that the two-temperature H$_2$O fits are tracing the average physical conditions representing radial excitation temperature and abundance gradients in each disk, which are expected to contribute to the rotational water spectrum given the large spread of upper-level energy levels spanned by the lines in MIRI spectra \citep{Banzatti_2023b}. Thus, we explore whether there is enough information in the MIRI-MRS spectra to retrieve the properties of these radial gradients, and whether straightforward parametric models can provide improved model fits to the data relative to using two-temperature models. The details are presented below.

\subsubsection{Power law Profiles}
\label{sec:powerlaw}

To fit a radial distribution of water vapor, the first necessary assumption we make is that the excitation temperature profiles can be described by a monotonically decreasing function of radius. We choose to approximate the underlying H$_2$O excitation temperature distribution using a power law \citep[e.g.][]{Kaeufer_2024} of the form

\begin{equation}
    T_{ex} = T_0 \left( \frac{r}{0.5 \text{ au}}\right)^{-\alpha}.
    \label{eq:tex_profile}
\end{equation}

Note that if the H$_2$O gas is thermalized with the dust, one would expect $a\sim 1/2$ for a passive irradiated disk \citep{Chiang_1997, Dullemond_2001}. For simplicity, we further assume that the distribution of water vapor is smooth and also described by a power law:

\begin{equation}
    N = N_0 \left( \frac{r}{0.5 \text{ au}}\right)^{-\beta},
\end{equation}

which could either increase or decrease with radius depending on dynamical, chemical, and/or optical depth effects. 

To implement these parametric models, we generate the H$_2$O spectra as the sum of 50 slabs representing concentric rings spanning from 0.1 to 10 au in each disk. The minimum radius of 0.1 au is an estimate based on the typical dust sublimation radius for the disks in our sample. We decide to construct the radial grid using uniformly $\log_{10}$-spaced points, in order to more finely sample the innermost 1 au region. Each ring has an emitting area given by:

\begin{equation}
    A = \pi (r_{out}^2 - r_{in}^2)  \cos{i},
\end{equation}

where $i$ is the disk inclination, and $r_{out}$ and $r_{in}$ are the outer and inner ring radii, respectively. The excitation temperature and column density of each ring is evaluated at the mid-point.

For these retrievals, we use the following Uniform prior distributions:

\begin{equation}
\begin{split}
\log_{10} T_{0} &= \mathcal{U} (2.0, 3.3) \text{ K}\\
\log_{10} N_{0} &= \mathcal{U} (12, 22) \text{ cm}^{-2}\\
\alpha &= \mathcal{U} (0.0, 5.0) \\
\beta &= \mathcal{U} (-5.0, 5.0), \\
\label{eq:uniform2}
\end{split}
\end{equation}

to allow the models a reasonably large amount of flexibility. Note that while these models consist of 50 slabs, they require fewer free parameters than the two-temperature fits (only four instead of six). 

\begin{figure}[ht!]
    \centering
    \includegraphics[width=0.45\textwidth]{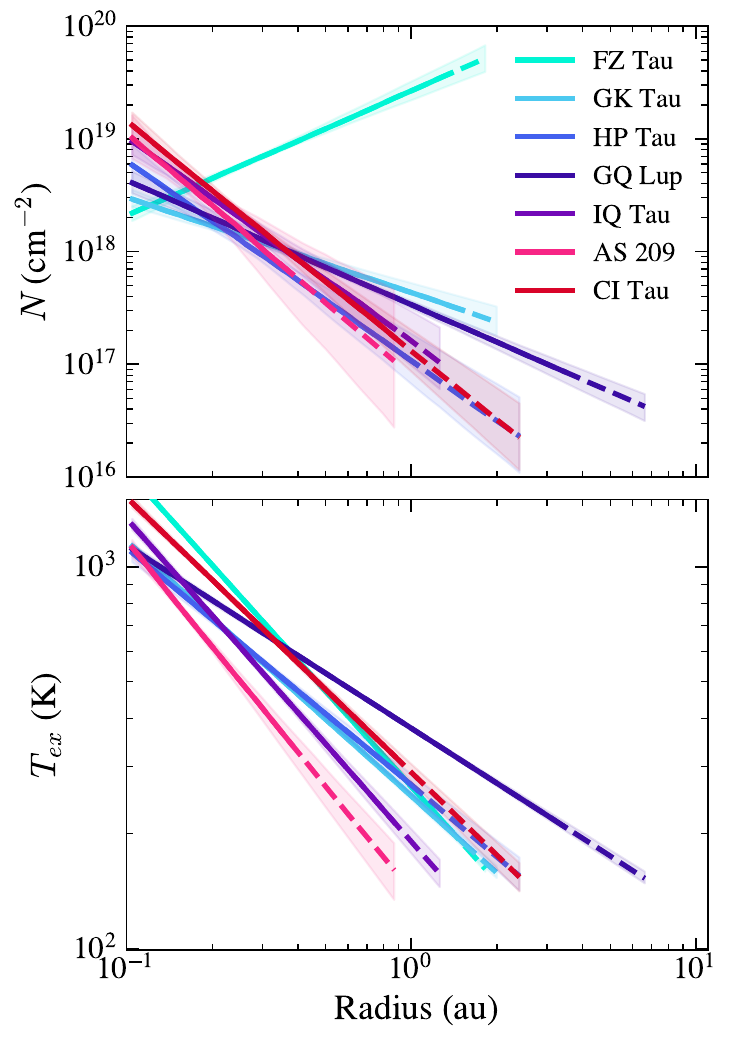}
    \caption{Power law H$_2$O column density (top) and excitation temperature (bottom) profiles fit to each disk in the sample. The solid lines correspond to the median profiles generated from posterior samples, while the shaded regions indicate the 0.16 and 0.84 percentiles at each radius. The profiles are truncated at 150 K. The dashed-line portions correspond to the regions where the modeled H$_2$O emission lies below the detection threshold (see Section \ref{sec:snowline?}).}
    \label{fig:powerlaw_profiles}
\end{figure}

The models achieve convergence for all disks in the sample, with statistically well-constrained parameters. Figure \ref{fig:powerlaw_profiles} shows the best-fit power law column density and excitation temperature profiles for each disk. The plotted profiles are truncated at the radius where $T_{ex}$ drops below 150 K, which we identify as the predicted surface snowline location. We find that extended disks exhibit steep column density profiles, with $\beta = 1.8 - 2.1$. On the other hand, compact disks exhibit a wide variety of column density profiles: the best-fit profiles for GK Tau and GQ Lup are similar, with $\beta = 0.9-1.0$, while that of HP Tau is much steeper ($\beta = 1.8$), akin to those of extended disks. Interestingly, FZ Tau exhibits a peculiar, radially-increasing H$_2$O column density profile, with $\beta = -1.1$, that peaks at $\sim 5 \times 10^{19}$ cm$^{-2}$ prior to freeze-out. As for the excitation temperature profiles, all disks exhibit profiles with $\alpha = 0.5 - 1.0$, steeper than expected for an optically-thin, radiation-dominated disk layer. We find an average $a$ of 0.73, with no clear temperature profile distinctions between compact and extended disks.

\subsubsection{Exponentially-tapered Column Density}
\label{sec:tapered}

One potential drawback of fitting power law column density profiles is that in order to reproduce the depletion of cold water vapor in extended disks---and the sharp depletion of water vapor at the snowline in any disk---the profiles may converge to steeper power law indices ($b$), which in turn increase the column density of hot water vapor. To explore whether the column densities of high-temperature H$_2$O found for extended disks are only artificially higher due to a restrictive parametric model, we next include an exponential taper to the column density profiles:

\begin{equation}
    N = N_0 \left( \frac{r}{0.5 \text{ au}}\right)^{-\beta} \exp \left[ - \left(\frac{r}{r_{\text{taper}}} \right)^{\phi} \right],
\end{equation}

where the taper radius $r_{\text{taper}}$ is a free parameter. Implementing the additional parameter $\phi$ thus allows us to model the depletion of cold water vapor without necessarily forcing the column density profile to increase in the innermost disk regions. For these fits, we maintain the Uniform priors described in Section \ref{sec:powerlaw} and use the following Uniform priors for $r_{\text{taper}}$ and $\phi$:

\begin{equation}
\begin{split}
\log_{10} r_{\text{taper}} &= \mathcal{U} (-1.0, 1.0), \\
\phi &= \mathcal{U} (0.0, 5.0).
\end{split}
\end{equation}

\begin{figure}[ht!]
    \centering
    \includegraphics[width=0.45\textwidth]{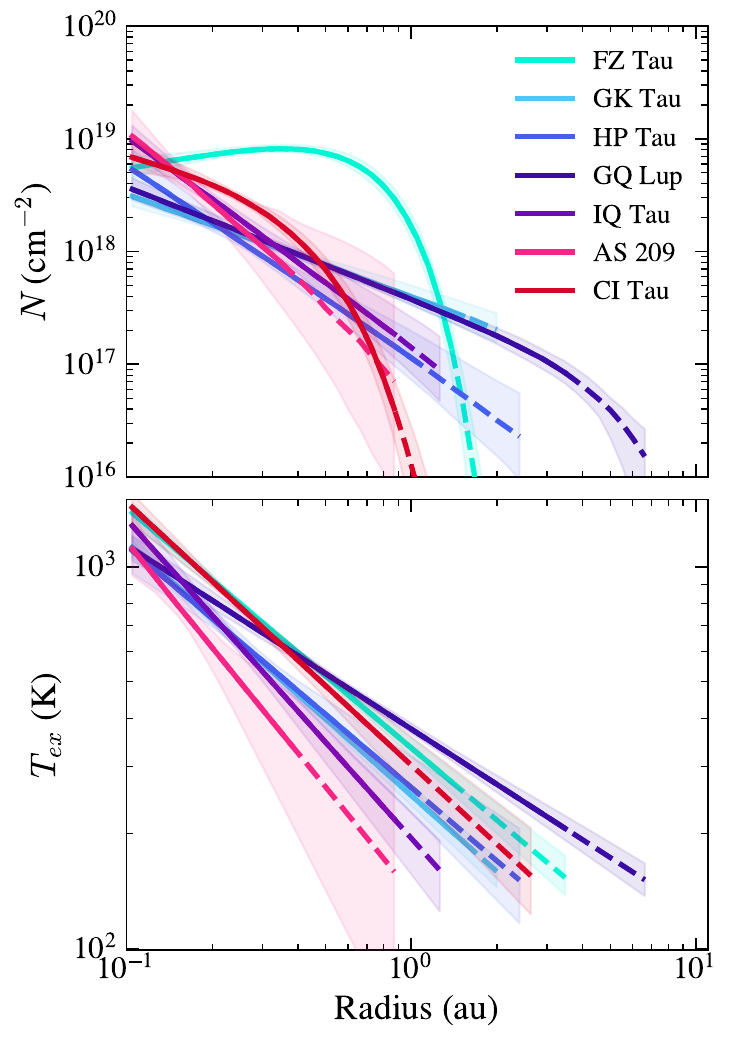}
    \caption{Same as Figure \ref{fig:powerlaw_profiles}, for the exponentially-tapered column density profile fits.}
    \label{fig:tapered_profiles}
\end{figure}

The best-fit profiles are presented in Figure \ref{fig:tapered_profiles}. Appendix \ref{apx:posteriors} shows example corner plots illustrating the posterior distributions obtained for FZ Tau. We find that including an exponential taper helps model the suppression of cold H$_2$O in extended disks while preventing the hot H$_2$O column densities from reaching unlikely high values ($>10^{19}$ cm$^{-2}$). The $N$ profiles of three compact disks, GK Tau, HP Tau, and GQ Lup converge to combinations of $\phi$ and $r_{\text{taper}}$ such that they only vary minimally compared to the previous simple power law profile fits. However, the column density profile of FZ Tau changed significantly. The profile now increases as $N \propto r^{0.5}$ out to a radius of around 0.7 au, beyond which $N$ drops steeply with $\phi \approx 2.4$. Note that in this case adding a taper has also prevented $N$ from reaching extremely high values out to the snowline. Finally, we do not observe significant changes to the excitation temperature profiles, except for that of FZ Tau, which is now less steep with $a$ down from 0.8 to 0.6.

\begin{figure*}[ht!]
    \centering
    \includegraphics[width=0.98\textwidth]{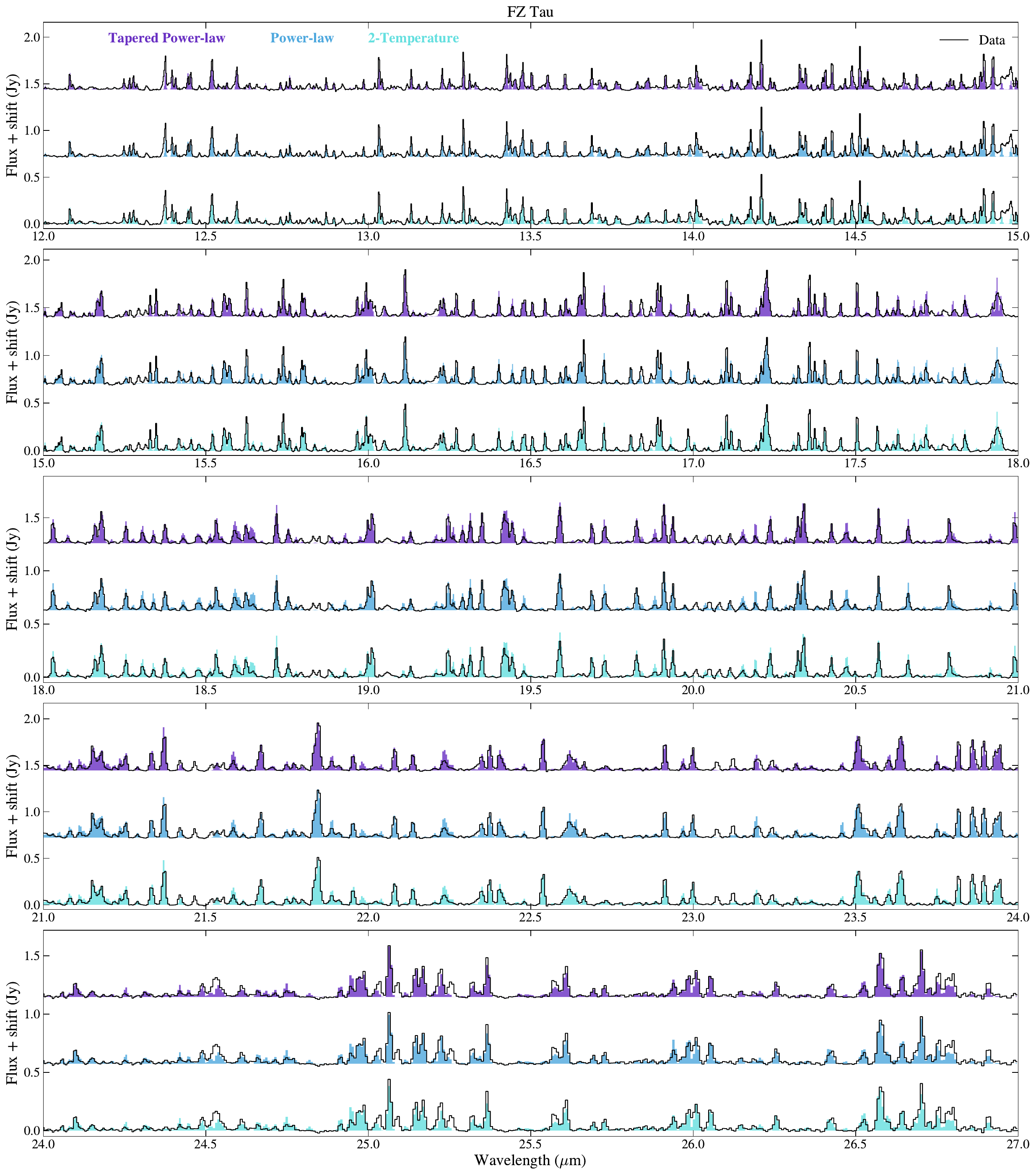}
    \caption{Best-fit H$_2$O emission models to the spectrum of the compact disk of FZ Tau. The two-temperature model is shown in cyan (bottom), the power law model in blue (middle), and the exponentially-tapered power law model in violet (top of each panel). The data is shown three times in each panel and the models are shifted vertically for clarity.}
    \label{fig:FZTau_compare_models}
\end{figure*}

In Figure \ref{fig:FZTau_compare_models}, we present a comparison between the two-temperature, power law, and exponentially-tapered power law best-fit models to FZ Tau. Note that all three prescriptions provide excellent fits to the data, and most differences are only noticed by close inspection of the residuals. For instance some line fluxes are under-predicted by the two-temperature model (see the region near 23.8 $\mu$m) and others are over-predicted (see 18.55, 19.6, 20.35 $\mu$m), while the tapered-power law model can better reproduce these lines. Distinctions like these are observed across the sample, but the emission lines mentioned above are not always affected in the same way. Similar comparisons are presented for each disk in Appendix \ref{apx:compare}. 

\begin{figure*}[ht!]
    \centering
    \includegraphics[width=0.9\textwidth]{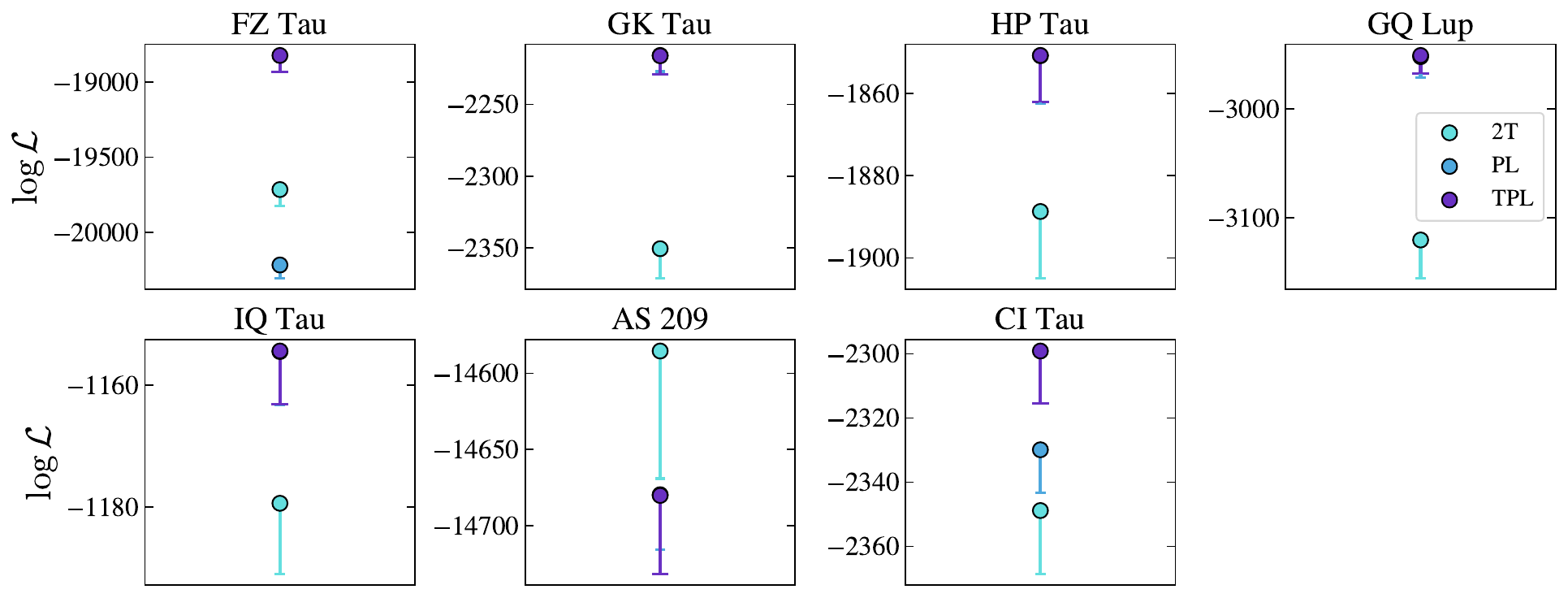}
    \caption{The maximum log likelihood (circles) attained for each H$_2$O emission model fit. The labels 2T, PL, and TPL correspond to the two-temperature, power law, and exponentially-tapered power law models, respectively. The vertical bars show the range of $\log \mathcal{L}$ values spanned by posterior distribution samples within the 0.025-0.975 percentiles. Note that fitting a temperature gradient always results in a statistically significant improvement, except in the case of AS 209.}
    \label{fig:compare_logl}
\end{figure*}

Figure \ref{fig:compare_logl} shows the maximum log-likelihood achieved with each model and the range of $\log \mathcal{L}$ spanned by the $0.025-0.975$ ($2\sigma$) percentiles of posterior samples for every disk in the sample. The percentiles provide one metric to evaluate which models provide better fits to each dataset. For FZ Tau, we note that a power law $N$ profile provides a worse fit compared to the two-temperature model, but a tapered power law then results in an improved fit. In general we find that fitting $T_{ex}$ and $N$ gradients results in better model fits relative to simply using two-temperature models, with $> 95\%$ confidence. This result is in agreement with \citet{Kaeufer_2024}, who also find statistically significant evidence for an H$_2$O column density and temperature gradient in the disk of GW Lup. Yet the specific parametric model used (power law versus tapered power law) only results in statistically significant fit improvements for FZ Tau and CI Tau. Still, the exponentially-tapered profiles are preferred over the power law profiles given that the latter occasionally produce unlikely high column density values\footnote{The line emission becomes highly optically thick at columns larger than 10$^{19}$.}. AS 209 is the only case in which both gradient fits perform significantly worse than the two-temperature model. This is further discussed in Section \ref{sec:dis}. Note that the comparison presented in Figure \ref{fig:compare_logl} is rather conservative, as both models with radial gradients not only achieve higher likelihoods, but also require the same or fewer free parameters. The detailed treatment of $T_{ex}$ and $N$ gradients is therefore preferred to accurately model water emission, especially in cases when the H$_2$O spectrum must be subtracted to identify and fit blended, less-abundant species.

\begin{figure*}[ht!]
    \centering
    \includegraphics[width=0.97\textwidth]{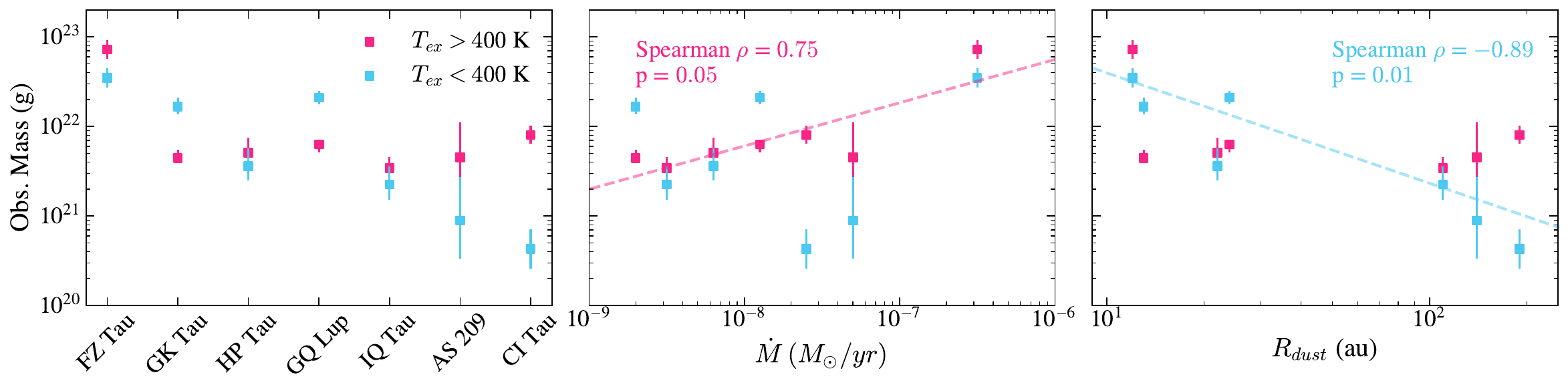}
    \caption{\textit{Left}: Observable cold ($150<T_{ex}<400$ K) and hot ($T_{ex}>400$ K) water vapor mass in each disk, calculated by integrating the best-fit exponentially-tapered power law column density profiles over each temperature range. \textit{Middle:} Observable H$_2$O masses as a function of stellar mass accretion rates. The Spearman correlation coefficient and p-value for the hot water mass is indicated on the top. The dashed line indicates a power law fit.\textit{Right}: Same as middle panel, but for H$_2$O masses vs. outer sub-mm dust disk radius. The correlation coefficient corresponds to the cold H$_2$O mass.}
    \label{fig:tapered_coldhotmass}
\end{figure*}


An additional benefit of fitting $N$ and $T_{ex}$ gradients is that we can now develop a well-defined measure of `cold' and `hot' water vapor in each disks. In particular, we define cold water vapor as that with $150 \leq T_{ex} < 400$ K, and hot water vapor that with $T_{ex} \geq 400$ K. The motivation behind the 400 K choice is twofold: First, 400 K accounts for radial and vertical diffusion of water vapor from the sublimation front. Second, 400 K is approximately 100 K above the temperature at which the gas-phase formation of water starts to become efficient \citep{Meijerink_2009}, thus ensuring that most of the water vapor classified as `hot' originates from gas-phase chemistry rather than sublimation. The observable cold and hot water vapor masses can then be found by integrating the column density profiles within the corresponding radial regions.  While the cold components may be contaminated by gas-phase formation of water, and the hot components by sublimation, we experimented using temperature cut-offs of up to 500 K and find no significant differences in the trends presented below.

\begin{deluxetable*}{lccccccccccl}
\label{tab:profile_results}
\tabletypesize{\small}
\tablecolumns{12}
\tablewidth{0pt}
\tablecaption{Radial H$_2$O profile fit results}
\tablehead{
\colhead{} & \multicolumn{4}{c}{Power law} & \colhead{} & \multicolumn{6}{c}{Tapered Power law} \\ \cline{2-5} \cline{7-12}
\colhead{Source} & \colhead{$N_{0}/10^{18}$} & \colhead{$T_{0}$} & \colhead{$\alpha$} & \colhead{$\beta$} & \colhead{} & \colhead{$N_{0}/10^{18}$} & \colhead{$T_{0}$} & \colhead{$\alpha$} & \colhead{$\beta$} &  \colhead{$\phi$} &  \colhead{$r_{taper}$}  \\
\colhead{} & \colhead{(cm$^{-2}$)} & \colhead{(K)} & \colhead{} & \colhead{} & \colhead{} & \colhead{(cm$^{-2}$)} & \colhead{(K)} & \colhead{} & \colhead{} &  \colhead{} &  \colhead{(au)} 
}
\startdata  
FZ Tau & $12.3^{+1.51}_{-1.11}$ & $472^{+4}_{-3}$ & $0.83^{+0.01}_{-0.01}$ & $-1.12^{+0.15}_{-0.13}$ & & $11.75^{+6.02}_{-2.58}$ & $518^{+5}_{-4}$& $0.63^{+0.02}_{-0.02}$ & $-0.47^{+0.26}_{-0.33}$ & $2.40^{+0.53}_{-0.55}$ & $0.71^{+0.12}_{-0.15}$ \\ 
GK Tau  & $0.78^{+0.09}_{-0.09}$ & $399^{+5}_{-5}$ & $0.67^{+0.02}_{-0.02}$ & $0.85^{+0.15}_{-0.16}$ & & $0.93^{+1.23}_{-0.19}$ & $399^{+5}_{-4}$& $0.66^{+0.02}_{-0.02}$ & $0.84^{+0.16}_{-0.20}$ & $1.18^{+1.72}_{-1.05}$ & $4.92^{+3.42}_{-4.54}$ \\ 
HP Tau  & $0.37^{+0.11}_{-0.09}$ & $412^{+15}_{-13}$ & $0.62^{+0.04}_{-0.05}$ & $1.78^{+0.34}_{-0.37}$ & & $0.86^{+0.44}_{-0.45}$  & $411^{+11}_{-14}$& $0.63^{+0.04}_{-0.04}$ & $1.56^{+0.39}_{-0.46}$ & $0.27^{+1.99}_{-0.19}$ & $1.45^{+3.72}_{-1.10}$ \\ 
GQ Lup & $0.72^{+0.06}_{-0.05}$ & $526^{+5}_{-4}$ & $0.48^{+0.01}_{-0.02}$  & $1.10^{+0.13}_{-0.11}$ & & $0.77^{+0.07}_{-0.06}$ & $525^{+5}_{-5}$& $0.48^{+0.01}_{-0.01}$ & $0.98^{+0.13}_{-0.12}$ & $3.30^{+1.18}_{-1.82}$ & $5.78^{+1.94}_{-1.48}$ \\ 
IQ Tau & $0.56^{+0.25}_{-0.15}$ & $344^{+13}_{-15}$ & $0.85^{+0.04}_{-0.04}$ & $1.82^{+0.32}_{-0.39}$ & & $0.57^{+0.26}_{-0.15}$ & $347^{+12}_{-14}$& $0.84^{+0.04}_{-0.04}$ & $1.80^{+0.38}_{-0.34}$ & $2.81^{+1.54}_{-1.41}$ & $3.93^{+3.16}_{-2.00}$ \\ 
AS 209  & $0.34^{+0.60}_{-0.21}$ & $267^{+30}_{-32}$ & $0.91^{+0.11}_{-0.11}$ & $2.14^{+0.85}_{-0.92}$ & & $0.43^{+1.54}_{-0.32}$ & $266^{+42}_{-42}$& $0.92^{+0.14}_{-0.13}$ & $2.09^{+1.15}_{-1.25}$ & $2.75^{+1.50}_{-1.69}$ & $1.85^{+3.64}_{-1.24}$ \\ 
CI Tau  & $0.54^{+0.13}_{-0.11}$ & $477^{+13}_{-13}$ & $0.72^{+0.04}_{-0.04}$ & $2.05^{+0.29}_{-0.32}$ & & $4.06^{+4.81}_{-2.47}$ & $488^{+17}_{-13}$& $0.69^{+0.03}_{-0.05}$ & $0.42^{+0.68}_{-0.51}$ & $1.62^{+0.84}_{-0.36}$ & $0.35^{+0.18}_{-0.10}$ \\ 
\enddata
\tablecomments{The reported values for each parameter consist of the median and 0.16-0.84 posterior distribution percentiles.}
\end{deluxetable*}

Figure \ref{fig:tapered_coldhotmass} shows the derived observable cold and hot H$_2$O masses for each disk, and as a function of mass accretion rate and outer sub-mm dust disk radius. These masses were calculated using the exponentially-tapered column density profiles. A Spearman correlation test reveals a statistically significant anti-correlation between cold water vapor mass and $R_{dust}$, and a tentative correlation between hot water vapor mass and the stellar mass accretion rate. The former exhibits a correlation coefficient $\rho = -0.89$ with $p$-value 0.01, and the latter has $\rho = 0.75$ with $p = 0.05$. In fact, the observed relations can be well-fit by power law expressions of the form
\begin{equation}
\begin{split}
    \log_{10} \left( \frac{M[\text{cold H}_2\text{O}]}{\text{g}} \right) &= p \log_{10} \left(\frac{R_{dust}}{\text{au}} \right) + q, \\
    \log_{10} \left( \frac{M[\text{hot H}_2\text{O}]}{\text{g}} \right) &= r \log_{10} \left(\frac{\dot{M}}{M_{\odot/yr}} \right)+ s,
\end{split}
\end{equation}
with $p = -1.23 \pm0.2$, $q = 23.8 \pm 0.4$, $r = 0.48 \pm0.1$, and $s = 25.6 \pm0.9$ as shown by the dashed lines in Figure \ref{fig:tapered_coldhotmass}. We consider the latter correlation tentative given the larger p-value, and because the high accretion rate of FZ Tau has a strong effect in driving the correlation. No significant correlation is found between cold water vapor mass and mass accretion rate, nor between hot water vapor mass and $R_{dust}$, even after dividing out the correlation with $\dot{M}$. Finally, to complement Figure \ref{fig:tapered_coldhotmass}, Figure \ref{fig:tapered_coldhotNA} in Appendix \ref{apx:extra_corr} shows the average column density ($\hat{N}$) and total emitting area for cold and hot water vapor in each disk.

\section{Interpretation and Discussion}  \label{sec:dis}


\subsection{Is the Snowline Detected?}
\label{sec:snowline?}
 We have carried out detailed retrievals to estimate the excitation conditions of water vapor in the innermost regions of seven protoplanetary disks, including the effects of radial gradients in excitation temperature and column density. As mentioned in Section \ref{sec:powerlaw}, the profiles are truncated at the location where $T_{ex}$ drops below 150 K, approximately the location of the expected surface H$_2$O snowline. However, we further examine whether or not H$_2$O emission down to 150 K is in fact detected by MIRI in each disk. To do this, we calculate the total wavelength-integrated flux contributed by each of the 50 slab models fit to each disk, for both the power law and tapered power law profiles. We then compare these to the estimated background noise (see Appendix \ref{apx:noise}), and find the radial/temperature bin at which the modeled integrated H$_2$O emission drops below a $5\sigma$ detection level. We denote these as emission cutoff radii $(R_{out})$ and cutoff excitation temperature $(T_{out})$.  We interpret these radii as the outermost mid-IR H$_2$O emission location in each disk.

\begin{figure*}[ht!]
    \centering
    \includegraphics[width=0.95\textwidth]{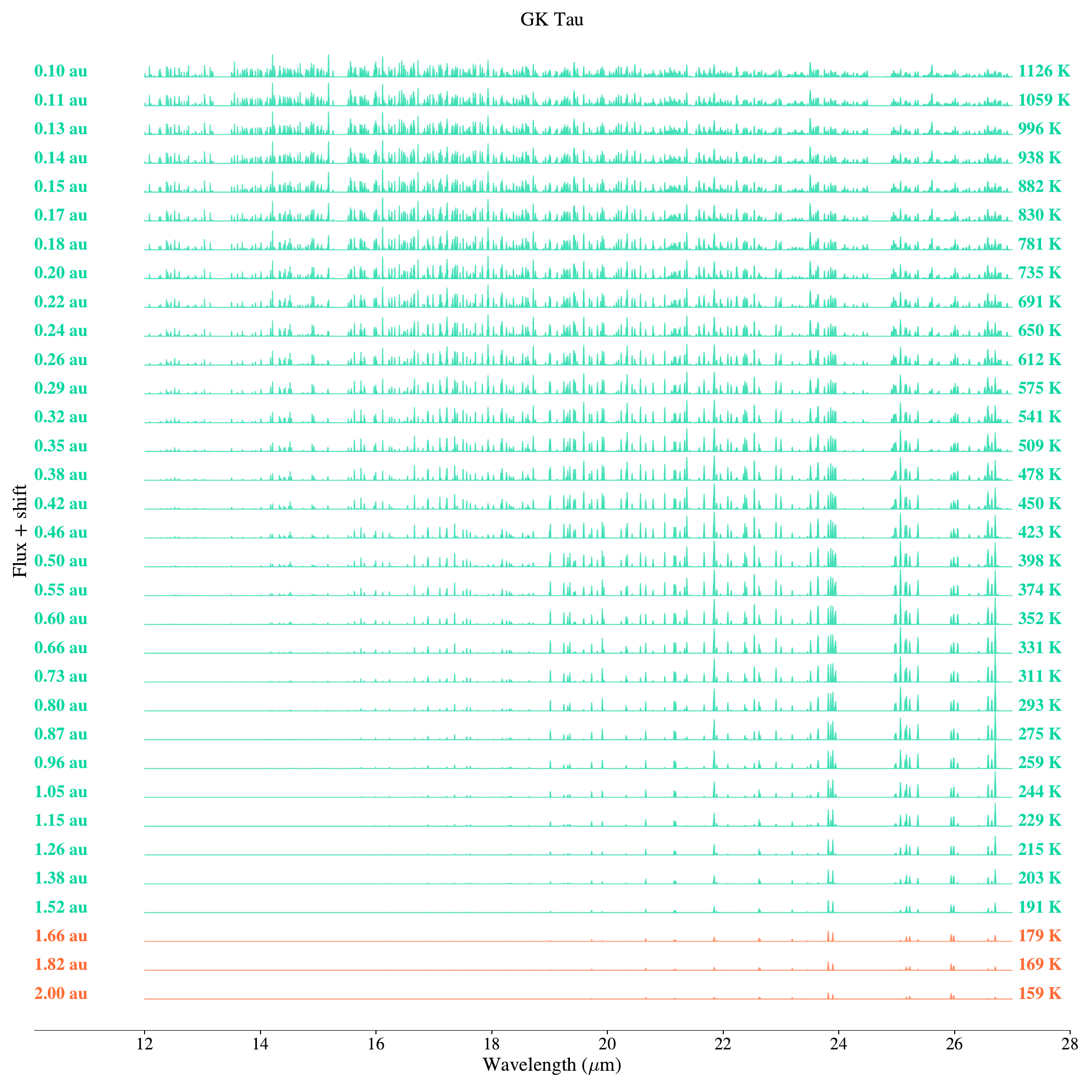}
    \caption{The contributions to the H$_2$O spectrum fit to the GK Tau data at each radial bin, for the exponentially-tapered column density model. The excitation temperature of each component is indicated on the right side. Models in green have integrated fluxes $>5 \sigma$, whereas the models in orange lie below the detection threshold. In this case, we detect H$_2$O emission with $T_{ex}$ down to $\sim180$ K.}
    \label{fig:GKTau_radial_spectra}
\end{figure*}

 Figure \ref{fig:GKTau_radial_spectra} shows the best-fit exponentially-tapered power law H$_2$O model for the disk of GK Tau, with the individual contributions of each slab spectrum shown individually. The models above the detection threshold are shown in green, while those undetected are shown in orange. In this case, $R_{out} = 1.7$ au, and emission down to $\sim180$ K is detected, which lies near the expected range for H$_2$O freeze-out. We detect H$_2$O emission with $T_{out}$ down to $\sim200$ K in two other disks: GQ Lup and IQ Tau. For the rest, the mid-IR cutoff occurs at temperatures $\gtrsim 250$ K. The precise mid-IR cutoff radius and temperature for each disk are reported in Table \ref{tab:outer_radii}. Note that the profiles shown in Figures \ref{fig:powerlaw_profiles} and \ref{fig:tapered_profiles} also indicate the location of $R_{out}$ in each disk.

  Our findings indicate that there are multiple H$_2$O emission lines within the MIRI-MRS wavelength coverage that are sensitive to gas with temperatures $<200$ K. However, these lines are weak, and are only detected in a few particularly water-rich disks. In the rest of the sample, the abundance of observable water vapor with $T_{ex}<200$ K is too low to be detected in the mid-IR. Thus, the reported location of the 150 K surface H$_2$O snowline in each disk should be considered a prediction, and its properties must be confirmed in the future by far-IR observations sensitive to the lowest-energy H$_2$O lines.

  We emphasize that truncating the column density profiles at either 150 K or $T_{out}$ has no significant effect on the cold and hot water vapor mass-radius correlations reported in Section \ref{sec:tapered}. Hence, whether a) the best-fit $N$ profiles can be extrapolated to $T_{ex}<T_{out}$ or b) the H$_2$O column density in this region drops due to e.g. chemical quenching \citep{Blevins_2016} or vertical diffusion \citep{Meijerink_2009} does not affect our conclusions.

\begin{figure}[ht!]
    \centering
    \includegraphics[width=0.45\textwidth]{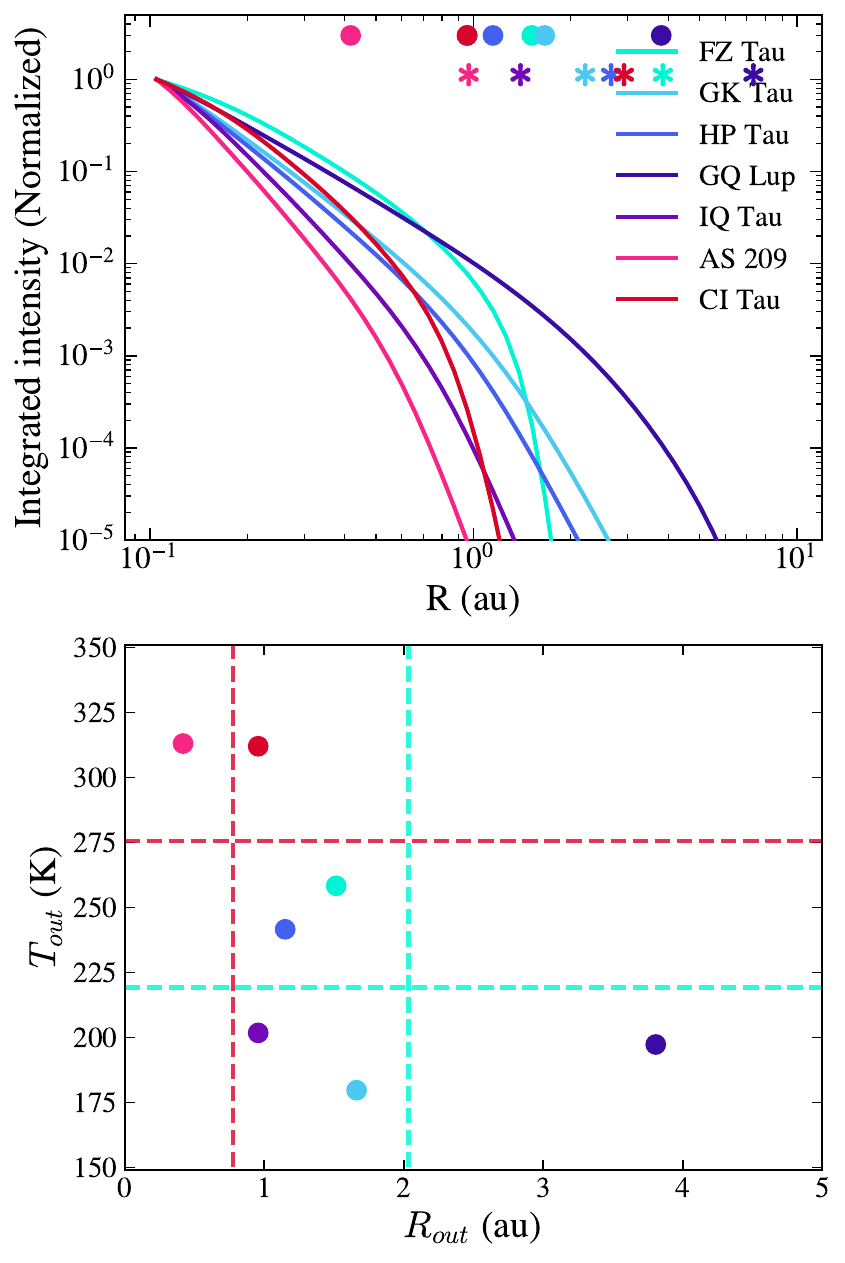}
    \caption{\textit{Top:} Normalized wavelength-integrated H$_2$O flux profiles as a function of radius, extracted from the best-fit tapered $N$ profiles. Asterisks indicate the radii where $T_{ex}$ reaches 150 K in each disk. Circles show the cutoff radius at which the integrated emission drops below 5$\sigma$. \textit{Bottom:} The H$_2$O cutoff radius for each disk, and the H$_2$O excitation temperature derived at that radius. Dashed lines indicate the average values for compact (blue) and extended (red) disks.}
    \label{fig:h2o_radii}
\end{figure}

  To complement Figure \ref{fig:GKTau_radial_spectra}, Figure \ref{fig:h2o_radii} shows the integrated H$_2$O fluxes as a function of radius for each disk, as well as $R_{out}$, $T_{out}$, and the predicted location of the surface snowline. The interpretation of $R_{out}$ is currently open. One possibility is that the mid-IR emission cutoff radius is an indicator of the \textit{midplane} snowline, which could be the case under the vertical cold-finger scenario \citep[e.g.][]{Meijerink_2009, Krijt_2016} and thus the ratio of $R_{out}$ to the $T_{ex} = 150$ K radius may provide a measure of vertical transport of gas in a disk. However, the derived $R_{out}$ are inconsistent with the expected location of the midplane snowlines based on viscous heating \citep{Min_2011, Mulders_2015}, and are further uncorrelated to $\dot{M}$, contrary to what one would expect \citep{Garaud_2007}. 

Rather, we find a notable distinction between the $R_{out}$ of compact and extended disks. As presented in the bottom panel of Figure \ref{fig:h2o_radii}, the mean H$_2$O cutoff radii for extended disks lies at around 0.8 au, whereas that of compact disks is located closer to 2 au. The vertical axis shows the excitation temperature reached at $R_{out}$, illustrating that, on average, most of the H$_2$O flux in compact disks originates from gas with $T_{ex} \gtrsim 220$ K, while in extended disks most of the flux comes from water vapor with  $T_{ex} \sim 50$ K warmer. $R_{out}$ thus appears to be tracing the zone where gas-phase formation of H$_2$O becomes efficient at $\gtrsim 300$ K in extended disks, whereas in compact disks $R_{out}$ may trace the radius where the column of H$_2$O sublimated from icy pebbles becomes detectable in the mid-IR (but not necessarily the theoretical location of the midplane snowline).

\subsection{Comparison with the Dust Temperature}
\label{sec:dust_temp}
The derived $T_{ex}$ profiles can be compared to the expected temperature structure in each disk. In particular, we compare the retrieved H$_2$O temperature profiles to a simple prescription for the dust temperature above the photosphere—from where the inner-disk water vapor spectrum is expected to arise \citep[e.g.][]{Woitke_2018, Bosman_2022}:
\begin{equation} \label{eq:cg97}
    T_{dust} = \frac{1}{\epsilon^{1/4}}\left( \frac{R_{*}}{2 r} \right)^{1/2} T_{*},
\end{equation}
where $R_{*}$ and $T_{*}$ are the stellar radii and effective temperature, and $\epsilon$ is the dust emissivity in the mid-IR \citep{Chiang_1997, Dullemond_2001}. Following \citet{Pontoppidan_2024}, we assume $\epsilon = 0.17$ for the dust size distribution of a standard T Tauri disk, and also explore the effects of setting $\epsilon = 1.0$ for dust growth beyond a few $\mu$m and $\epsilon = 0.058$ for an ISM dust distribution.

\begin{figure*}[ht!]
    \centering
    \includegraphics[width=0.98\textwidth]{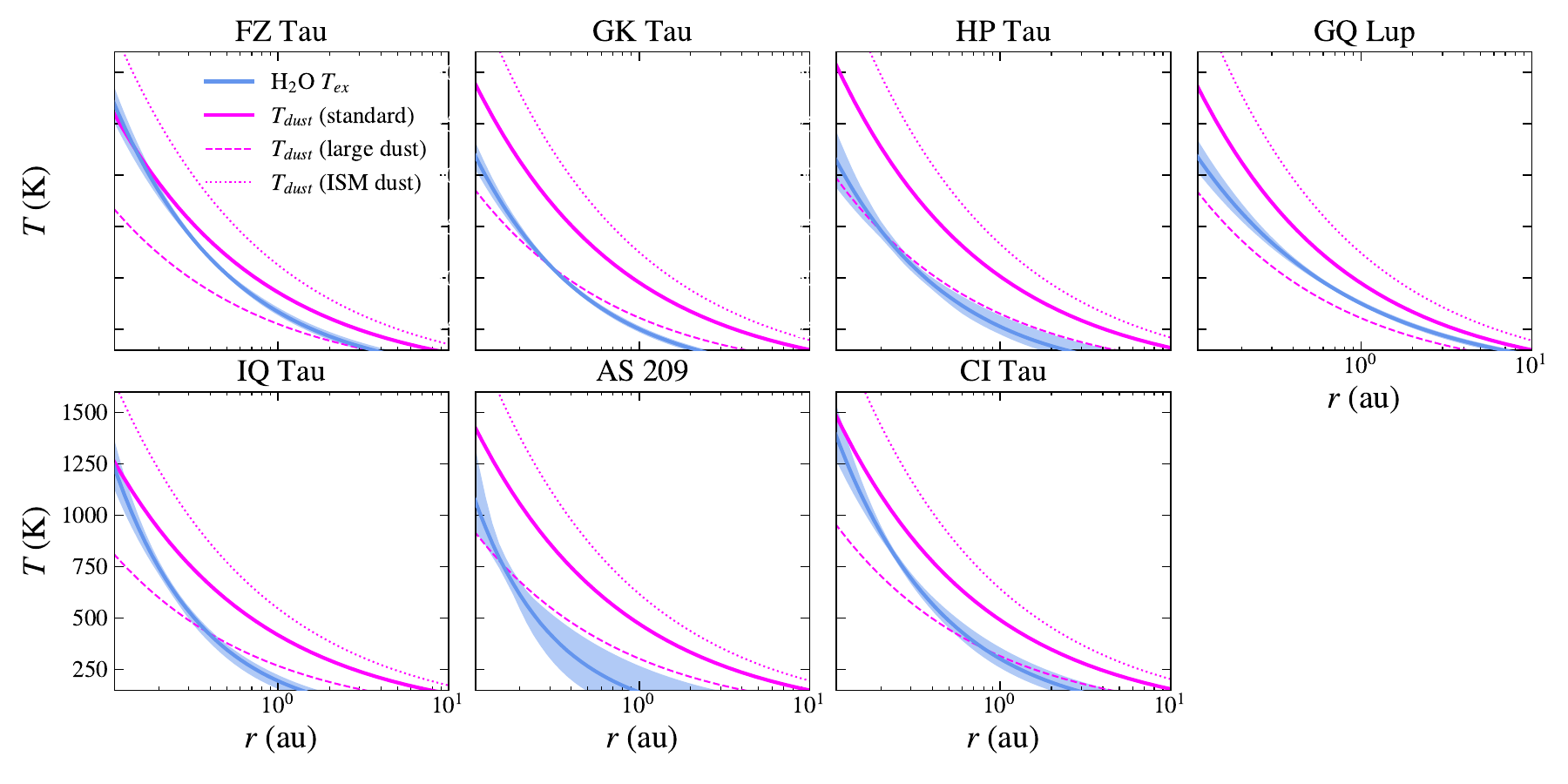}
    \caption{The excitation temperature profiles (blue) derived for H$_2$O in each disk with $1\sigma$ confidence regions, compared to the expected disk surface dust temperature (magenta) for different dust size distributions (see Section \ref{sec:dust_temp}).}
    \label{fig:Tex_profiles}
\end{figure*}

Figure \ref{fig:Tex_profiles} shows the modeled dust temperature profiles, compared to the H$_2$O excitation temperature profiles fit to each disk. Two results stand out: First, the standard dust temperature appears to be always moderately higher than the H$_2$O excitation temperature. Second, the H$_2$O $T_{ex}$ profiles are steeper than the dust temperature profiles. The only disk in which the retrieved H$_2$O $T_{ex}$ is in very good agreement with the dust temperature is FZ Tau, which is in line with the comparison reported by \citet{Pontoppidan_2024}.

Most thermochemical models predict that the gas and dust temperatures are decoupled in the uppermost disk layers. However, when decoupled, PAH photo-electric heating is predicted to raise the gas temperature by hundreds to thousands of K \citep[e.g.][]{Kamp_2004, Jonkheid_2004, Woitke_2009}. Evidently, we find no indication of H$_2$O gas super-heating. Yet it is unclear what mechanism would lead the gas to be colder than the dust. Assuming that the gas responsible for rotational H$_2$O emission is thermalized with H$_2$ and $T_{ex} \approx T_{gas}$, our results may rather simply indicate that gas and dust are coupled and Equation \ref{eq:cg97} is over-predicting the standard dust temperature. As suggested by Figure \ref{fig:Tex_profiles}, the dust temperatures could be lower and in better agreement with the typical H$_2$O $T_{ex}$ if there had been significant dust growth in the inner regions of these disks. The steepness of some profiles may further arise due to a radial gradient in dust sizes. Alternatively, the dust temperature could also be lower than expected due to geometric effects such as shadowing from the inner rim \citep{Dullemond_2001, Rafikov_2006}. Ultimately the MIRI continuum emission may be used for detailed modeling of the dust properties and temperature in each of these disks, but such analysis is beyond the scope of this paper.

The retrieved $T_{ex}$ profiles are generally well constrained, with power law indices $a\gtrsim 1/2$, steeper than expected for the dust based on radiative equilibrium alone. Interestingly, the average temperature power law index of $0.69$ found for our sample is in excellent agreement with that predicted by thermochemical models ($\sim 0.7$) including the effects of H$_2$O self-shielding and chemical heating from photo-dissociation and molecule formation \citep{Bosman_2022}. In this context, the H$_2$O gas could be cooler than the dust by a few hundred K if H$_2$O self-shielding is highly efficient and chemical heating processes are suppressed. Note that the average power law index is also in good agreement with that found for the surface temperature of the GW Lup disk (index of 0.63), based on an H$_2$O temperature gradient fit to the MIRI spectrum \citep{Kaeufer_2024}.

\begin{deluxetable}{lccl}
\label{tab:outer_radii}
\tabletypesize{\small}
\tablecolumns{4}
\tablewidth{0pt}
\tablecaption{Outer H$_2$O cut-off parameters}
\tablehead{
\colhead{Source} & \colhead{$R_{\text{150K}}$} &  \colhead{$R_{\text{out}}$} &  \colhead{$T_{\text{out}}$} \\
\colhead{} & \colhead{(au)} &  \colhead{(au)}  & \colhead{(K)} 
}
\startdata  
FZ Tau & 3.81 & 1.51 & 258 \\ 
GK Tau & 2.19 & 1.67 & 179 \\ 
HP Tau & 2.63 & 1.15 & 241 \\ 
GQ Lup & 7.25 & 3.81 & 197 \\ 
IQ Tau & 1.38 & 0.96 & 202 \\ 
AS 209 & 0.96 & 0.42 & 313 \\ 
CI Tau & 2.89 & 0.96 & 312 \\ 
\enddata
\tablecomments{\textit{From left to right:} The predicted surface snowline location based on the exponentially-tapered models, the outer emission cutt-off radius, and the corresponding H$_2$O excitation temperature at that radius.}
\end{deluxetable}

Viscous heating may also play some role in setting the H$_2$O temperature profiles. While heating from accretion is generally expected to be deposited near the midplane, most of the viscous heat could rather be released near the H$_2$O-emitting disk surface via the magnetorotational instability \citep{Hirose_2011, Adamkovics_2014}. A simple power law may then be insufficient to reproduce the underlying temperature profile, and a more flexible model may be necessary to account for viscous and radiative heating-dominated regions.

Finally, it is possible that viscous heating and dust growth/disk shadowing effects collude in such a way that the H$_2$O $T_{ex}$ profiles are in fact tracing the dust temperature. That is, the H$_2$O emitting region is dominated by a mix of viscous and radiative heating, and the dust temperature modeled by Equation \ref{eq:cg97} may be an over-estimate. This would explain the possible correlation between hot water vapor and mass accretion rates, the steep $T_{ex}$ profiles, and why FZ Tau---which has the highest $\dot{M}$---is the only disk where H$_2$O $T_{ex}$  and Equation \ref{eq:cg97} are in good agreement. 

\subsection{The Ring-Like Water Vapor Distribution of FZ Tau}

The best-fit tapered power law $N$ radial profile reveals an uncommon H$_2$O column density distribution in the disk of FZ Tau. As opposed to those from the rest of the sample, the $N$ profile in FZ Tau exhibits a ring-like distribution, with $N\propto r^{1/2}$ out to 0.7 au and an abrupt exponential drop-off that brings $N$ below $10^{16}$ cm$^{-2}$ well within the expected freeze-out location (based both on our fits and Herschel observations \citep{Blevins_2016}). It appears unlikely that this phenomenon is somehow the result of inner-disk dust substructure (i.e. a cavity), given the particularly high $\log_{10} \dot{M} = -6.5$ of FZ Tau and its abundant cold water vapor reservoir. 

Conversely, it may be that the high accretion rate of FZ Tau leads to the observed ring-like H$_2$O distribution. If the midplane temperature exhibits a large gradient and the radial velocity of icy pebbles is large enough, the bulk of the latter might drift past the location of the water snowline and sublimate in a noticeably warmer disk region \citep[e.g.][]{Piso_2015}. This could lead to the the paucity of water vapor with $T_{ex}<300$ K, yet still produce a significant enrichment in H$_2$O with $ T_{ex} > 400$ K. The small drop in H$_2$O column density closer to the star could then be explained by the very high temperatures produced by viscous heating leading to the thermal dissociation of some water vapor (or direct photo-dissociation from the radiation field produced by the accretion itself). Indeed, kinematically-resolved ro-vibrational lines have been observed at 5 $\mu$m with iSHELL from the ground \citep{Banzatti_2023a}, and found to have a FWHM of 40 km/s (corresponding to a Keplerian orbit of 0.2 au) and a temperature of 1000 K, matching well the hotter part of the temperature gradient estimated in this work. Furthermore, \citet{Salyk_2019} also find evidence of a ring-like water vapor distribution in FZ Tau based on TEXES observations of 12 $\mu$m H$_2$O line emission. In this case the rotational H$_2$O lines are consistent with an inner emission cut-off radius of 0.4 au, where we also find evidence of a drop in H$_2$O column density.

\subsection{The Temperature Structure of AS 209}
\label{sec:tempas209}
AS 209 is the disk where we observe the strongest discrepancy between the H$_2$O excitation temperature profile and the expected dust temperature from radiative equilibrium. The retrieved $T_{ex}$ profile is the steepest in our sample, and the excitation temperatures are always several hundred K lower than $T_{dust}$. As discussed above, the steep power law index closer to 1 may suggest that the gas temperature is dominated by viscous heating. However, this is difficult to reconcile with the very low temperatures, even if $T_{dust}$ from Equation \ref{eq:cg97} was over-estimated. 

Note that AS 209 is the only disk for which the two-temperature H$_2$O model fit statistically outperforms both $T_{ex}$ gradient fits, and the cold H$_2$O component has a much lower temperature ($\sim$250 K) than those found in the rest of the sample. Thus, a more speculative possibility is that the water vapor structure in AS 209 is indeed not well described by a continuous radial distribution: A gas gap somewhere within the innermost 1 au could explain why the disk of AS 209 appears to be depleted in water vapor with $T_{ex}$ within $400-600$ K, but still exhibits a typical hot (800 K) water vapor reservoir and cool 250-300 K component that could be tracing the region beyond the proposed gap. Recent high-resolution iSHELL observations to ro-vibrational CO in AS 209 indeed support the presence of an inner gap from the significant discontinuity in velocity between a broad and narrow component of CO (Figure 18 in \citet{Banzatti_2023a}). Using the line velocity, we can estimate that a narrow gap would be somewhere between 0.16 au (from the peak of the broad component) and 0.56 au (from the width of the narrow component). The narrow CO component is also cold (possibly as cold as the cold H$_2$O in MIRI), as it significantly drops already at $J\sim25$ while in other disks it is usually strong up to higher $J$ levels. Still, the AS 209 MIRI spectrum has the lowest SNR and line-to-continuum ratio of those included in this study, and it is entirely possible the model mismatch is a side-effect of the low data quality.

\subsection{Icy Pebble Drift}

We have followed-up on the analysis carried out in \citet{Banzatti_2023b}, and confirmed that the trend between low-energy H$_2$O line fluxes and $R_{dust}$ corresponds to a significant anti-correlation between the latter and the observable water vapor mass with $150 < T_{ex} < 400 $ K. This new result supports the proposed scenario where compact disks have more efficient icy pebble drift, and thus present an increased flux of H$_2$O sublimation near the snowline. By contrast the hot water vapor \texttt{mass} does not correlate with the dust disk size, confirming previous results based on line fluxes alone \citep{Banzatti_2023b}.

Assuming that the cold water reservoir in each disk is dominated by H$_2$O sublimation from ices, we can obtain a simple estimate of the pebble mass flux into the inner disk from the observable water vapor mass as 
\begin{equation}
\label{eq:pebflux}
    \dot{M}_{peb} \approx \left(\frac{M[\text{Cold H}_2\text{O}]}{t_{\text{H}_2\text{O}}}\right)  \left(\frac{\eta}{f_{ice} f_{\text{H}_2\text{O}}}\right),
\end{equation}
where $t_{\text{H}_2\text{O}}$ is the approximate time that sublimated H$_2$O resides in the gas phase, $\eta$ is a correction factor to account for the unobservable water column under the optically thick layer, $f_{ice}$ is the ice fraction of drifting pebbles, and $f_{\text{H}_2\text{O}}$ is the fraction of molecular ice in water. Based on thermochemical modeling, $\eta$ is expected to be of order $10^{3}$ for water vapor with temperatures of 150-400 K \citep{Meijerink_2009, Bosman_2022}. We set $\eta = 10^3$. For $f_{ice}$ and $f_{\text{H}_2\text{O}}$ we assume 0.20 and 0.80, which are typically assumed values \citep[e.g.][]{Johansen_2021}. Finally, we assume $t_{\text{H}_2\text{O}}$ is set by the combination of diffusion $(t_{\text{diff}})$ and chemical $t_{\text{chem}}$ timescales. For the former, we can estimate the time it takes for water vapor sublimated from ice in the midplane to reach a typical emitting region at one scale height,
\begin{equation}
\label{eq:diff}
    t_{\text{diff}} = \frac{H}{\alpha c_s},
\end{equation}
where $c_s$ is the sound speed \citep[e.g.][]{Semenov_2011}

For a typical midplane snowline at 1 au and gas temperature of 200 K, $t_{\text{diff}}$ can range between 15-150 yr depending on the assumed inner disk viscosity $(\alpha = 0.01-0.001)$. Once in an elevated disk region, thermochemical models suggest water vapor will only reside in the gas phase for a $t_{\text{chem}}$ of 10-50 yr (Calahan, J. priv. comm.) before being photo-dissociated. We hence proceed with an assumed $t_{\text{H}_2\text{O}}$ of 100 yr. 

\begin{deluxetable}{lccc}
\label{tab:fluxes}
\tabletypesize{\small}
\tablecolumns{4}
\tablewidth{0pt}
\tablecaption{Pebble flux estimates}
\tablehead{
\colhead{Source} & \colhead{$t_{\text{disk}}$} & \colhead{$\dot{M}_{peb}$} & \colhead{$M_{peb}$}\\
\colhead{} & \colhead{$(Myr)$} & \colhead{($M_{\oplus}/Myr$)} & \colhead{($M_{\oplus}$)} 
}
\startdata 
FZ Tau & $1.1$ & $370$ & $410$ \\
GK Tau & $1.2$ & $220$ & $260$\\
HP Tau & $2.4$ & $70$ & $170$\\
GQ Lup & $1.0$ & $330$ & $330$\\
IQ Tau & $4.2$ & $30$ & $130$\\
AS 209 & $1.5$ & $30$ & $45$\\
CI Tau & $2.5$ & $<10$ & $10$\\
\enddata
\tablecomments{Pebble fluxes estimated from Equation \ref{eq:pebflux}. Disk ages obtained from \citet{neuhauser_2005, Andrews_2018, Long_2019, McClure_2019}.}
\end{deluxetable}

Table \ref{tab:fluxes} provides the assumed disk ages and estimates of the pebble mass flux and total pebble mass deposited in the inner disk $(M_{peb})$. While extended disks have pebble fluxes of a few tens of earth masses per Myr, those of compact disks are an order of magnitude larger. The compact disk of HP Tau appears to be an exception, with a noticeably lower cold water vapor mass and diminished $\dot{M}_{peb}$ of $70$ $M_{\oplus}/Myr$. As discussed in \citet{Banzatti_2023b}, the near infrared index of HP Tau shows tentative evidence of a dust cavity within the innermost 2 au. As such, HP Tau may have a different physical structure where molecular gas is affected by different irradiation conditions within a dust cavity, and thus exhibits a reduced cold H$_2$O reservoir that mimics those of extended disks. 

The disk of GQ Lup presents another interesting case. Despite being the largest of the `compact' disks, it exhibits a pebble mass flux comparable to that of the smallest disk in the sample, FZ Tau. As shown in the rightmost panel in Figure \ref{fig:tapered_coldhotmass}, its cold water vapor mass appears to be higher than expected from the $R_{dust}$ anti-correlation inferred from the rest of the sample. Furthermore, its $R_{out}$ is a factor $\sim 2$ greater than those of the other compact disks. These anomalies may be explained by the binary nature of GQ Lup, which is known to have a close brown-dwarf-mass companion at $\sim 110$ au \citep{neuhauser_2005}: both dynamical simulations \citep{Zagaria_2021a} and observations \citep{Zagaria_2021b} suggest that radial dust drift is substantially faster in binary systems due to tidal interactions, which would explain the higher abundance of cold water and yet a regular hot H$_2$O reservoir. The binary interactions may also lead to higher gas temperatures than expected for an isolated disk  \citep{Muller2012, Picogna_2013}, which could lead to the particularly extended outer H$_2$O radii found for GQ Lup.

The pebble mass flux estimates hint that compact and extended disks not only have distinct inner disk compositions, but may also result in the formation of different exoplanet populations. For instance, \citet{Lambrechts_2019} predict that disks may form either small terrestrial or super-Earth planets depending whether the total deposited pebble mass towards the inner disk is $\lesssim 100$ or $\gtrsim 200 M_{\oplus}$, respectively. In this framework, the compact disks in our sample may be forming systems of water-rich super-Earths, while the large disks may only form small rocky planets \citep[see also discussion in][]{vanderMarel_2021}. Note that there is substantial uncertainty regarding the parameters in Equation \ref{eq:pebflux} and hence on the exact pebble flux values. Thus, the key conclusion is that our analysis provides initial evidence that compact disks experience approximately an order of magnitude higher pebble fluxes compared to extended disks. Very high resolution sub-mm observations \citep[e.g.][]{Guerra_2024}, together with source-specific thermochemical modeling will be necessary to obtain more accurate inner-disk pebble flux constraints, and refine the connection between the latter and the observable cold H$_2$O reservoir.

\section{Summary}  \label{sec:conc}

The main results of this work are summarized below:

\begin{enumerate}
    \item We fit two-temperature water vapor slab models to the MIRI-MRS spectra of four compact and three extended disks around T Tauri stars. In general, the emission is very well-characterized by the mixture of an extended (1-10 au$^2$) 400 K component and a compact (0.1 au$^2$) 800 K component.
    
    \item We infer the spatial distribution of water vapor by fitting parametric radial excitation temperature and column density profiles. These models achieve even smaller residuals compared to the two-temperature approximation. Thus, when possible, we recommend this kind of modeling to accurately subtract water vapor emission and search for less abundant species in mid-IR disk spectra.
    
    \item We find that the observable cold water vapor mass is anti-correlated with the sub-mm dust disk size, confirming previously reported line flux correlations. The hot water vapor mass is instead tentatively correlated with stellar mass accretion rate, which may set the mid-IR emitting area.  
    
    \item The water vapor excitation temperature profiles appear to be steeper and colder than expected for the `super-heated' dust temperature above the disk photosphere for a passive irradiated disk. The average power law index of 0.7 is in agreement with thermochemical model predictions. 
    
    \item The disk of AS 209 is the only one in our sample that can be better modeled by two independent temperature components rather than a continuous H$_2$O distribution. This could be due to inner disk substructure.
    
    \item The disk of FZ Tau exhibits a ring-like column density radial profile, which may be explained by its steep radial temperature gradient and high mass accretion rate.
    
    \item Our models indicate that the bulk of the water vapor emits from a smaller and warmer radial region in extended disks compared to compact disks. This difference in emitting regions is not correlated to the mass accretion rate, and supports the icy pebble drift hypothesis.

    \item We estimate icy pebble mass fluxes from the observable cold water masses. Compact disks have typical fluxes an order of magnitude higher than those of extended disks. This may suggest compact disks are relatively more likely to form systems of super-earths.

\end{enumerate}

\section*{Acknowledgments}
The JWST data presented in this article were obtained from the Mikulski Archive for Space Telescopes (MAST) at the Space Telescope Science Institute. The specific observations analyzed can be accessed via \dataset[doi: 10.17909/q60w-vr09]{https://doi.org/10.17909/q60w-vr09}.

K.I.Ö. acknowledges support from the Simons Foundation (SCOL \#321183), an award from the Simons Foundation (\#321183FY19), and an NSF AAG Grant (\#1907653).

A portion of this research was carried out at the Jet Propulsion Laboratory, California Institute of Technology, under a contract with the National Aeronautics and Space Administration (80NM0018D0004).

\vspace{5mm}
\facilities{JWST (MIRI-MRS)}

\software{ iris \citep{Romero_Mirza_InfraRed_Isothermal_2023}, tinyGP \citep{Foreman-Mackey_2024}, dynesty \citep{Speagle_2020}, astropy \citep{astropy_1, astropy_2, astropy_3}, JAX \citep{Bradbury_2018}, matplotlib \citep{Caswell_2024}
          }

\appendix

\section{Full Spectra}
\label{apx:full}

\begin{figure*}[ht!]
    \centering
    \includegraphics[width=0.98\textwidth]{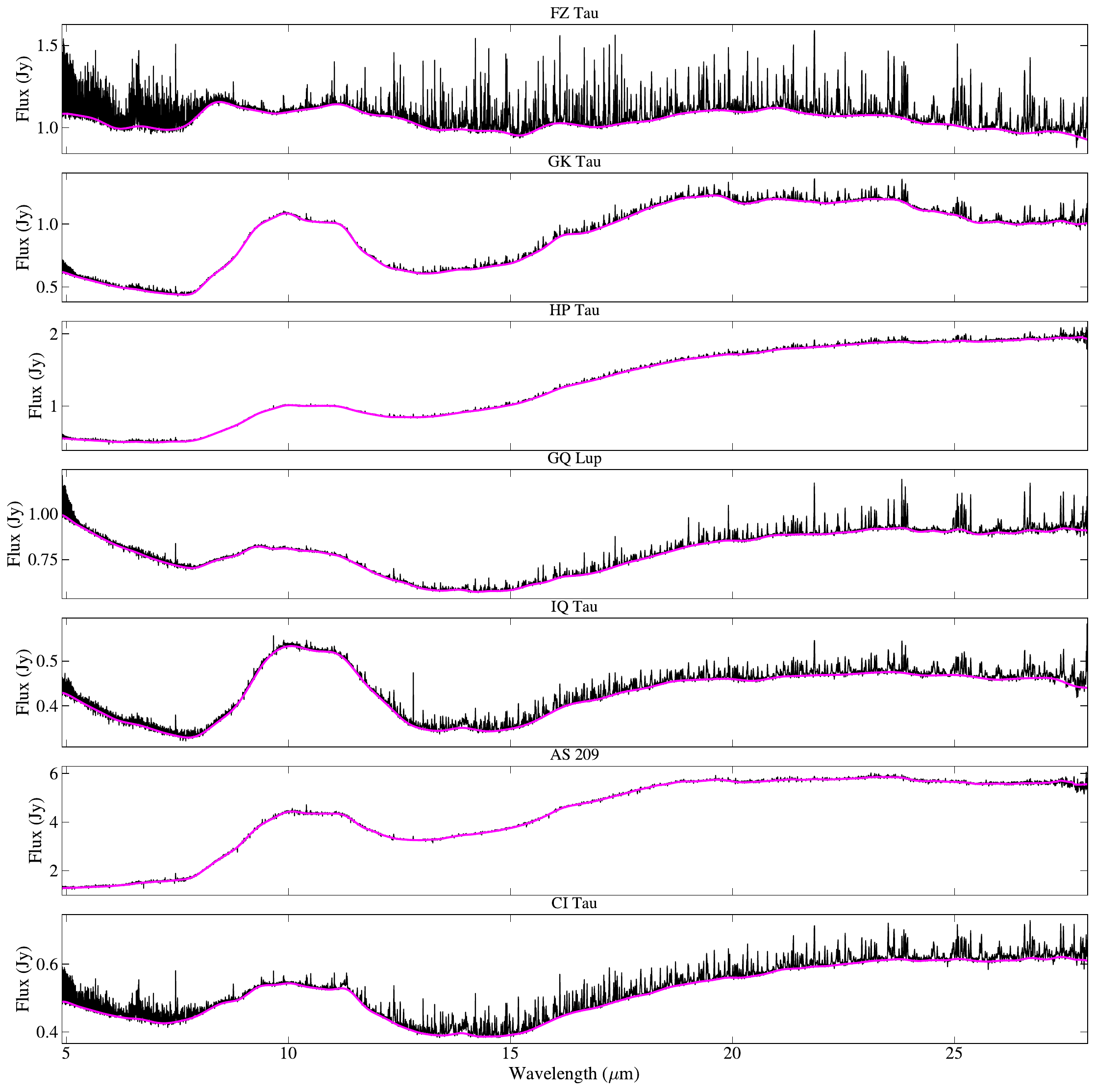}
    \caption{The full MIRI-MRS spectrum of each disk (black), and the corresponding continuum emission fit (magenta).}
    \label{fig:full_spectra}
\end{figure*}

Figure \ref{fig:full_spectra} presents the full MIRI-MRS spectra and best-fit continuum models. The continuum emission in each disk is fit following the technique described in \citet{Romero_Mirza_2024}.

\section{Gaussian Noise Estimation}
\label{apx:noise}

\begin{figure*}[ht!]
    \centering
    \includegraphics[width=0.98\textwidth]{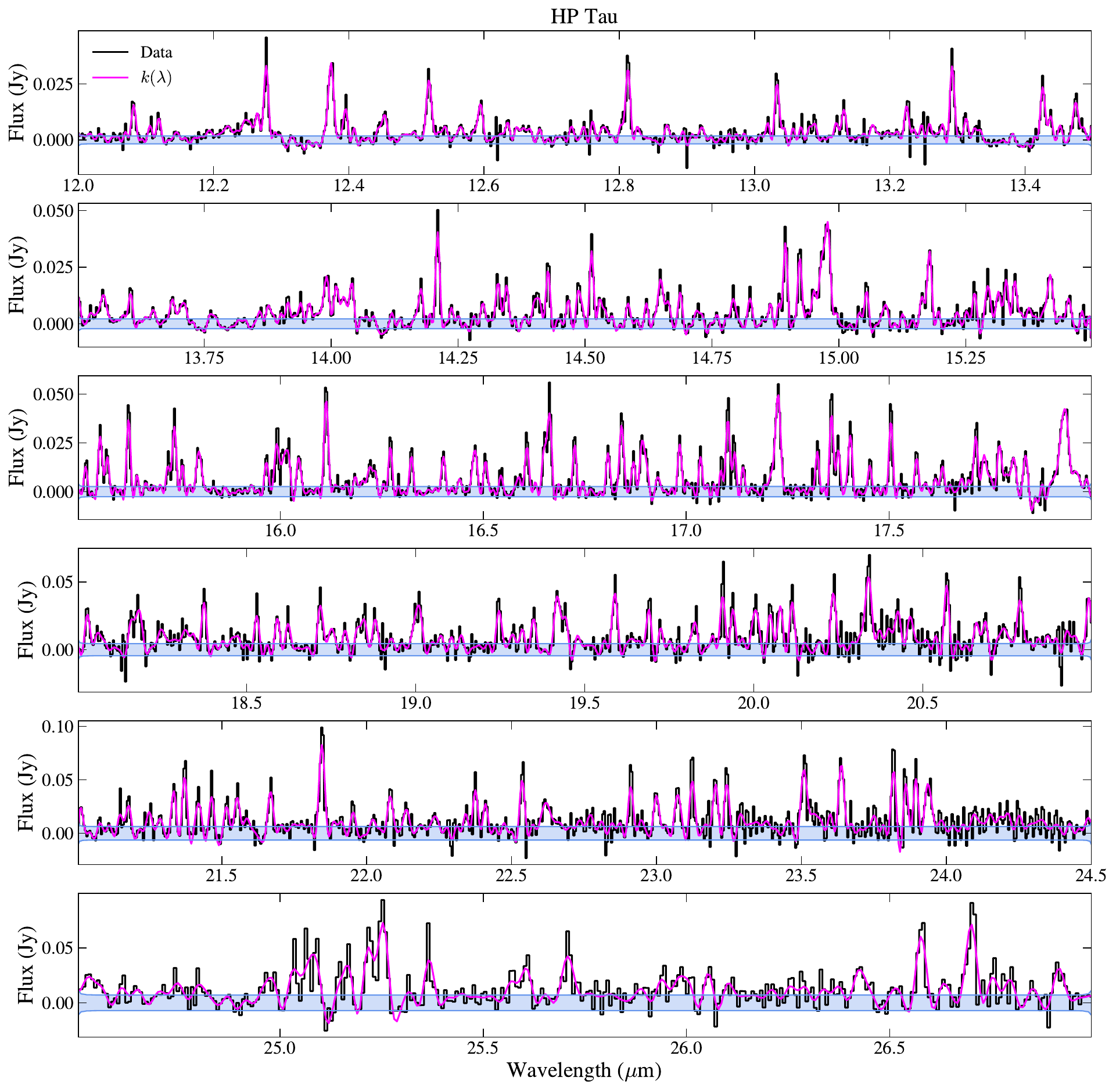}
    \caption{Gaussian Process decomposition of the spectrum of HP Tau. The black line shows the MIRI-MRS continuum-subtracted spectrum, the magenta line shows the conditioned squared exponential kernel, and the blue shaded region the best-fit $\pm 1 \sigma$ region. The latter includes the uncertainty associated with the continuum baseline fit.}
    \label{fig:HPTau_noise}
\end{figure*}

We use a Gaussian Process to estimate the amplitude of the background Normally-distributed noise in the continuum-subtracted spectra. In particular, we model the data in each of the six wavelength regions used in our analysis using a squared exponential covariance function $k(\lambda)$ with fixed amplitude 10.0 mJy \citep[e.g.][]{Romero_Mirza_2024} and a correlation length scale given by $2 \times \lambda_m/R$ (see Table \ref{tab:resolving_power}). We then condition this model on the data, and fit for the variance $\sigma^2$. That is, we assume any spectral features with correlation length smaller than the estimated resolving power in each region arise due to Gaussian noise. Inevitably, this results in some real spectral features being classified as background noise, and thus a rather conservative estimate of the flux uncertainty. Figure \ref{fig:HPTau_noise} illustrates the resulting decomposition in the case of HP Tau. Finally, we add a constant 10.0 mJy uncertainty in quadrature to the best-fit $\sigma$, which accounts for the typical uncertainty in the continuum baseline fit. This sum is the uncertainty used to calculate the Gaussian log-likelihood in the Bayesian retrievals described in Section \ref{sec:mod_and_res}. The noise estimates are listed in Table \ref{tab:noises}.

\begin{deluxetable}{lcccccc}[ht!]
\label{tab:noises}
\tabletypesize{\footnotesize}
\tablecolumns{7}
\tablewidth{0pt}
\tablecaption{Gaussian noise}
\tablehead{
\colhead{Source} & \colhead{$\sigma [12.0 -13.5] \mu$m} & \colhead{$\sigma [13.5-15.5] \mu$m} & \colhead{$\sigma [15.5 - 18.0] \mu$m} & \colhead{$\sigma [18.0 - 21.0] \mu$m} & \colhead{$\sigma [21.0 - 24.5] \mu$m} & \colhead{$\sigma [24.5 - 27.0] \mu$m} \\
\colhead{} & \colhead{(Jy)} & \colhead{(Jy)} & \colhead{(Jy)} & \colhead{(Jy)} & \colhead{(Jy)} & \colhead{(Jy)}
}
\startdata 
FZ Tau & 0.026 & 0.031 & 0.044 & 0.035 & 0.033 & 0.035 \\
GK Tau & 0.010 & 0.010 & 0.012 & 0.014 & 0.015 & 0.015 \\
HP Tau & 0.010 & 0.010 & 0.011 & 0.013 & 0.015 & 0.014 \\
GQ Lup & 0.010 & 0.010 & 0.012 & 0.014 & 0.017 & 0.019 \\
IQ Tau & 0.010 & 0.010 & 0.010 & 0.011 & 0.011 & 0.011 \\
AS 209 & 0.016 & 0.014 & 0.018 & 0.025 & 0.031 & 0.032 \\
CI Tau & 0.011 & 0.011 & 0.012 & 0.012 & 0.012 & 0.013 \\
\enddata
\tablecomments{The best-fit Gaussian noise $(1\sigma)$ to each wavelength region included in this study.}
\end{deluxetable}

\section{Two-temperature Fits}
\label{apx:2t}

\begin{figure*}[ht!]
    \centering
    \includegraphics[width=0.98\textwidth]{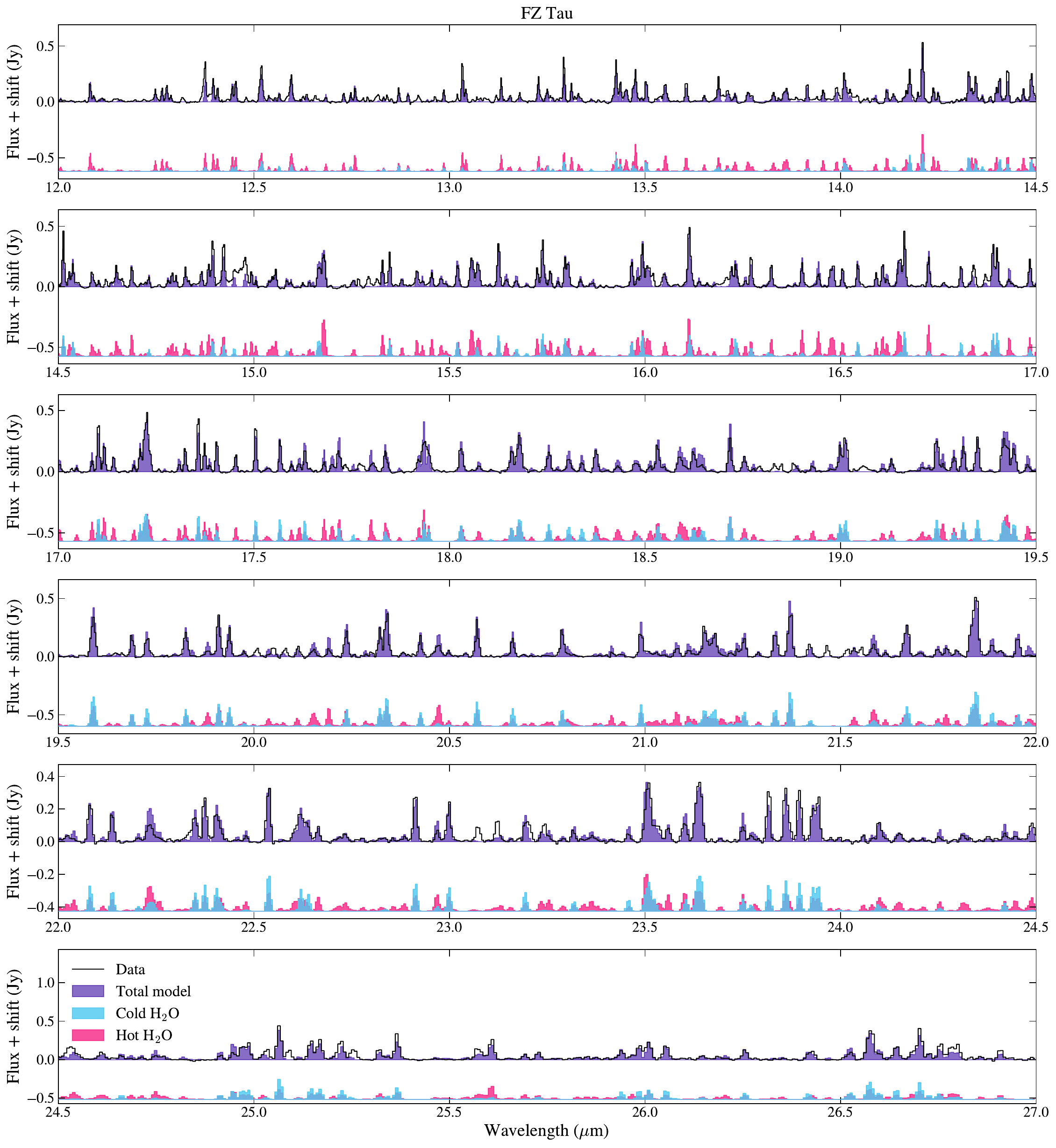}
    \caption{Same as Figure \ref{fig:gqlup_2comp}, for the compact disk of FZ Tau.}
    \label{fig:FZTau_2comp}
\end{figure*}

\begin{figure*}[ht!]
    \centering
    \includegraphics[width=0.98\textwidth]{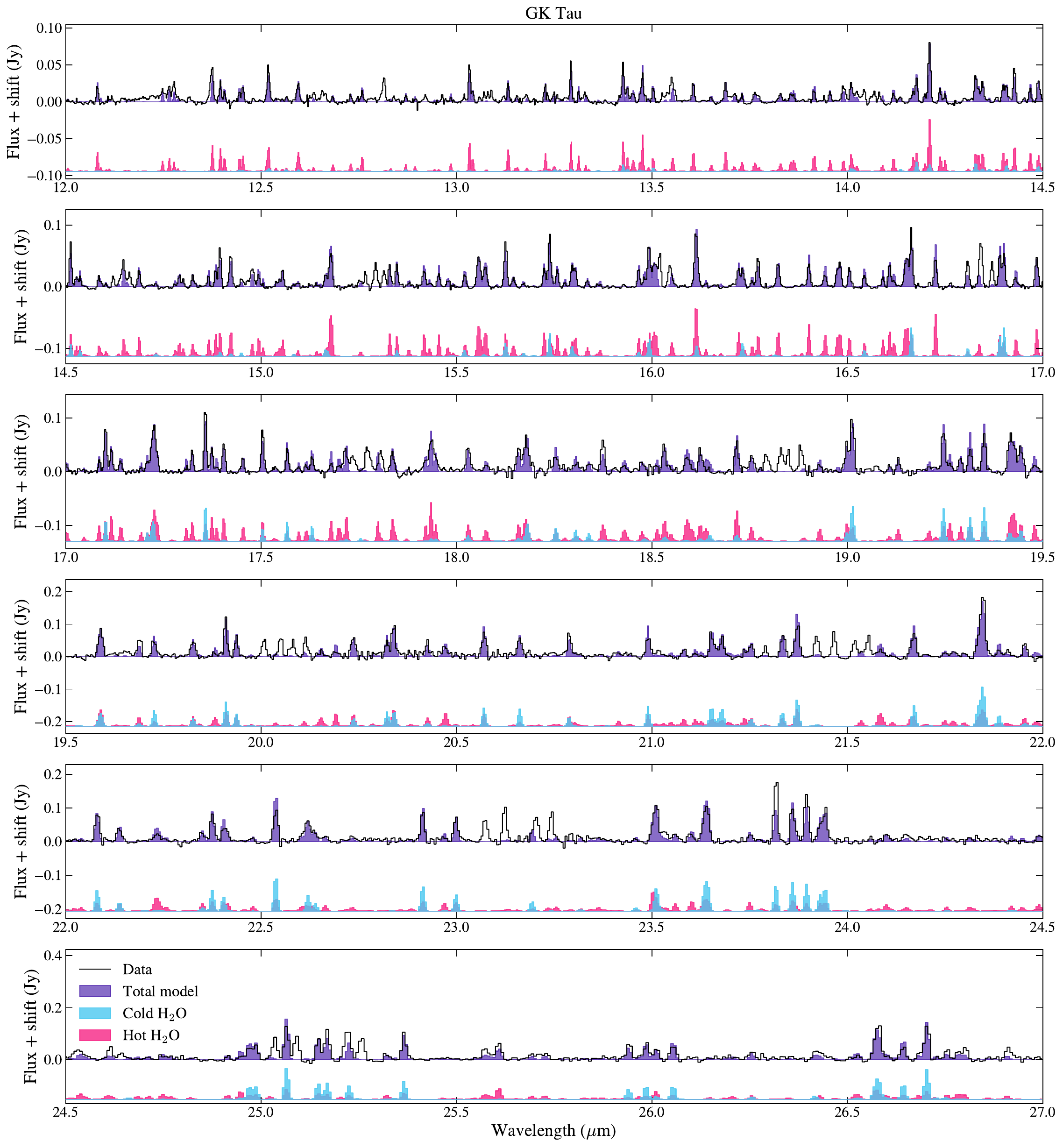}
    \caption{Same as Figure \ref{fig:gqlup_2comp}, for the compact disk of GK Tau.}
    \label{fig:GKTau_2comp}
\end{figure*}

\begin{figure*}[ht!]
    \centering
    \includegraphics[width=0.98\textwidth]{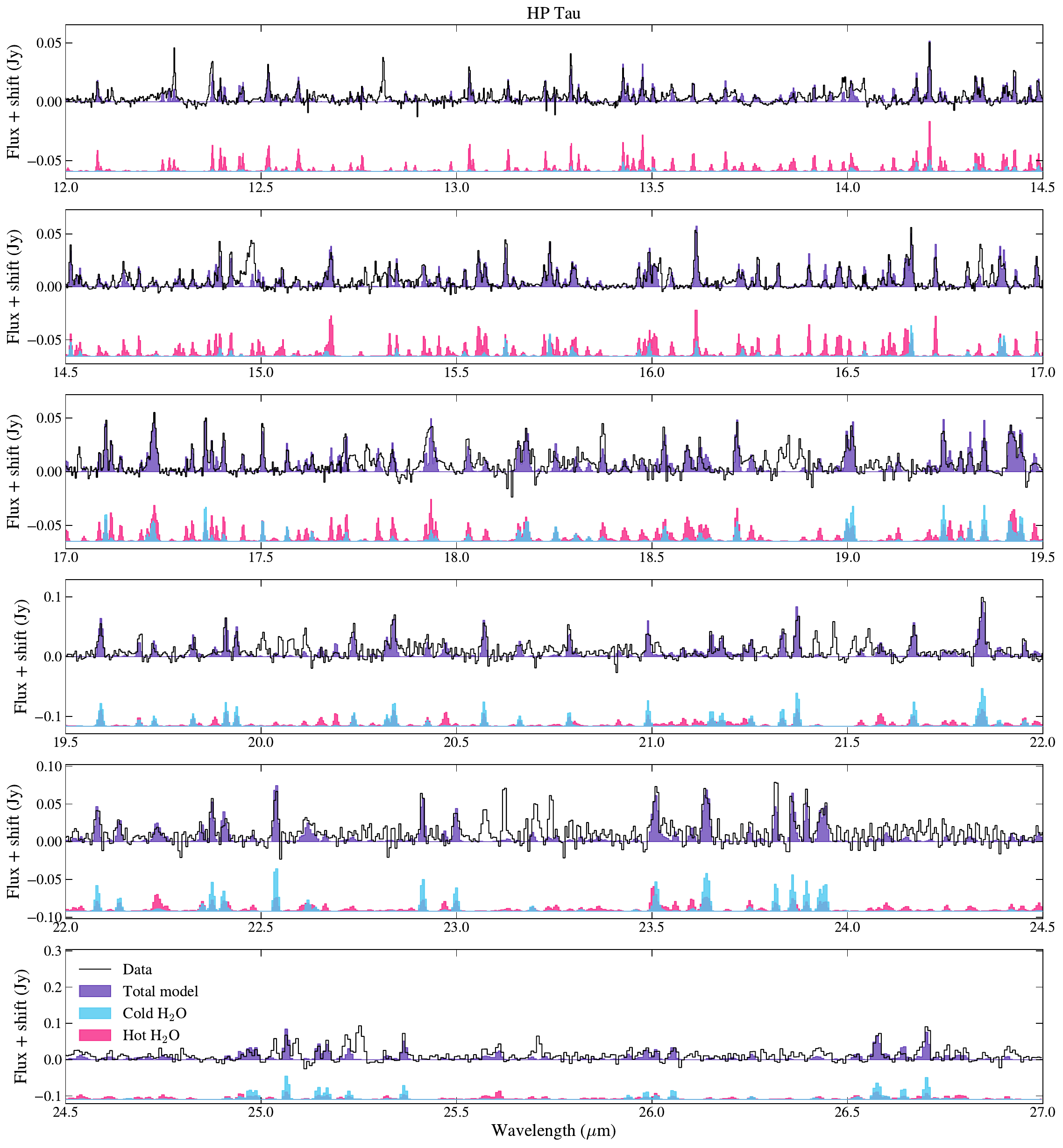}
    \caption{Same as Figure \ref{fig:gqlup_2comp}, for the compact disk of HP Tau.}
    \label{fig:HPTau_2comp}
\end{figure*}

\begin{figure*}[ht!]
    \centering
    \includegraphics[width=0.98\textwidth]{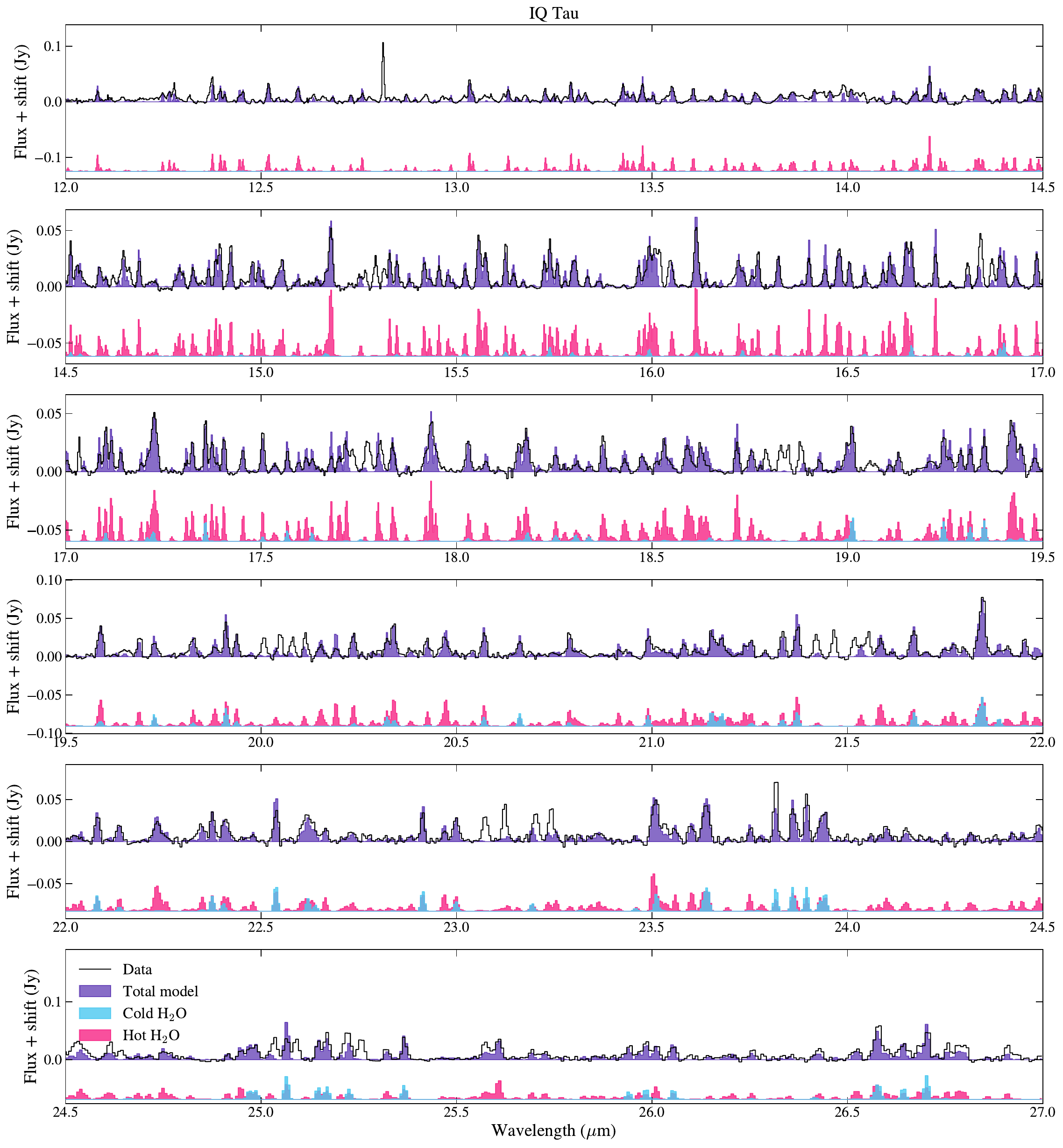}
    \caption{Same as Figure \ref{fig:gqlup_2comp}, for the extended disk of IQ Tau.}
    \label{fig:IQTau_2comp}
\end{figure*}

\begin{figure*}[ht!]
    \centering
    \includegraphics[width=0.98\textwidth]{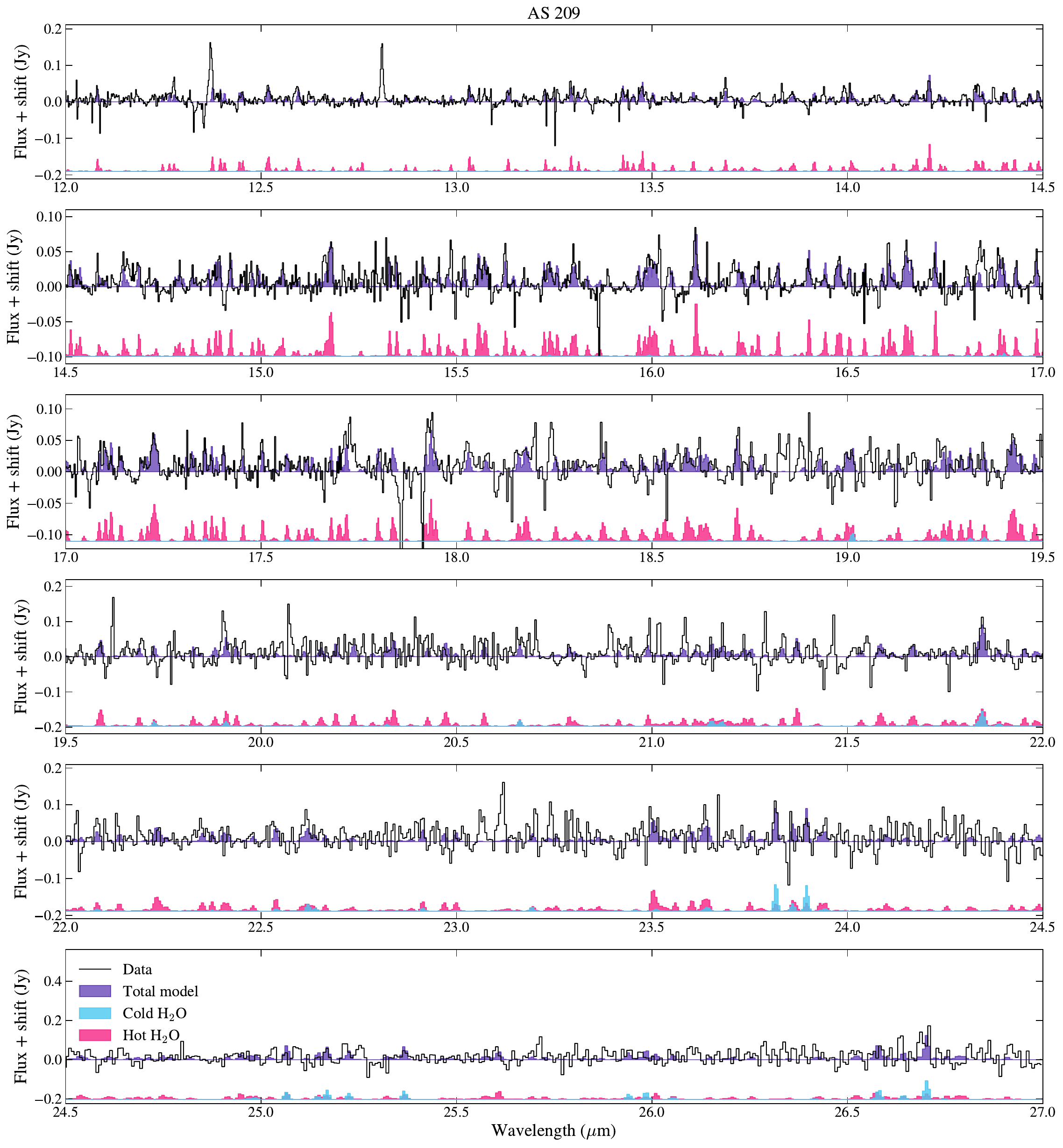}
    \caption{Same as Figure \ref{fig:gqlup_2comp}, for the extended disk of AS 209.}
    \label{fig:AS209_2comp}
\end{figure*}

In Figures \ref{fig:FZTau_2comp}-\ref{fig:AS209_2comp} we present the best-fit two-temperature H$_2$O models for the four disks not shown in Section \ref{sec:mod_and_res}, showing the individual contributions from hot and cold water vapor. 

\section{Model Comparison}
\label{apx:compare}

\begin{figure*}[ht!]
    \centering
    \includegraphics[width=0.98\textwidth]{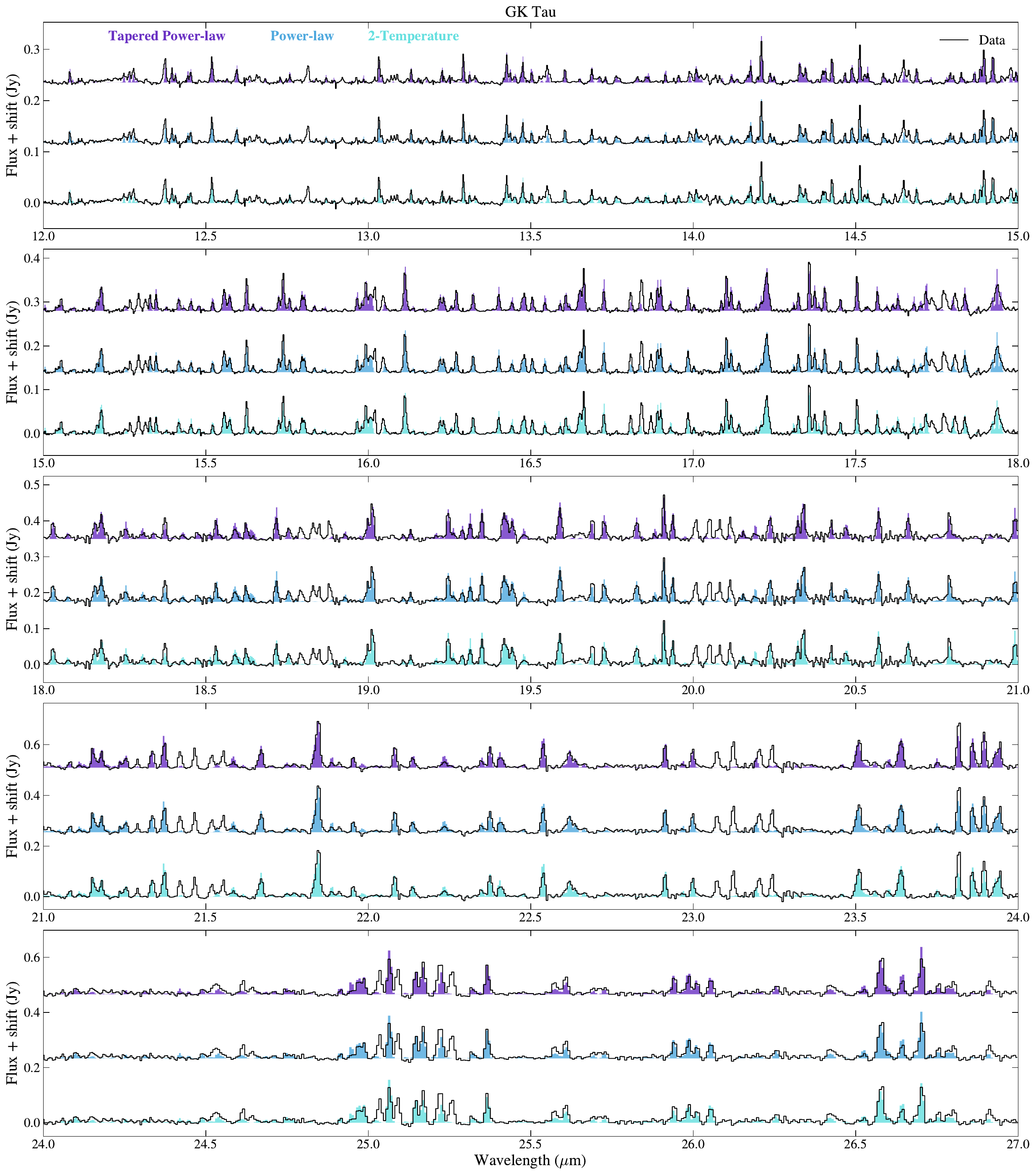}
    \caption{Same as Figure \ref{fig:FZTau_compare_models}, for GK Tau.}
    \label{fig:GKTau_compare_models}
\end{figure*}

\begin{figure*}[ht!]
    \centering
    \includegraphics[width=0.98\textwidth]{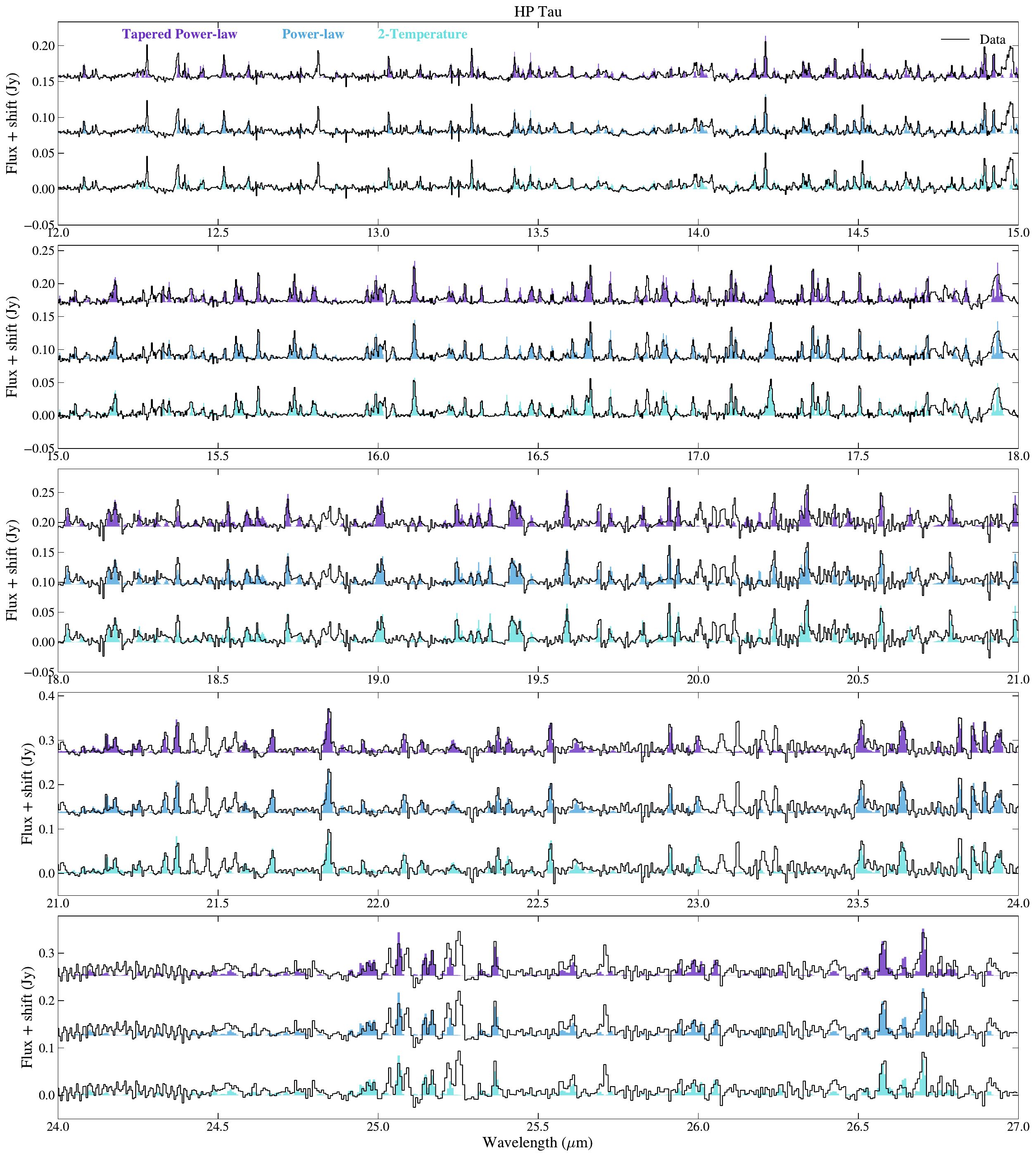}
    \caption{Same as Figure \ref{fig:FZTau_compare_models}, for HP Tau.}
    \label{fig:HPTau_compare_models}
\end{figure*}

\begin{figure*}[ht!]
    \centering
    \includegraphics[width=0.98\textwidth]{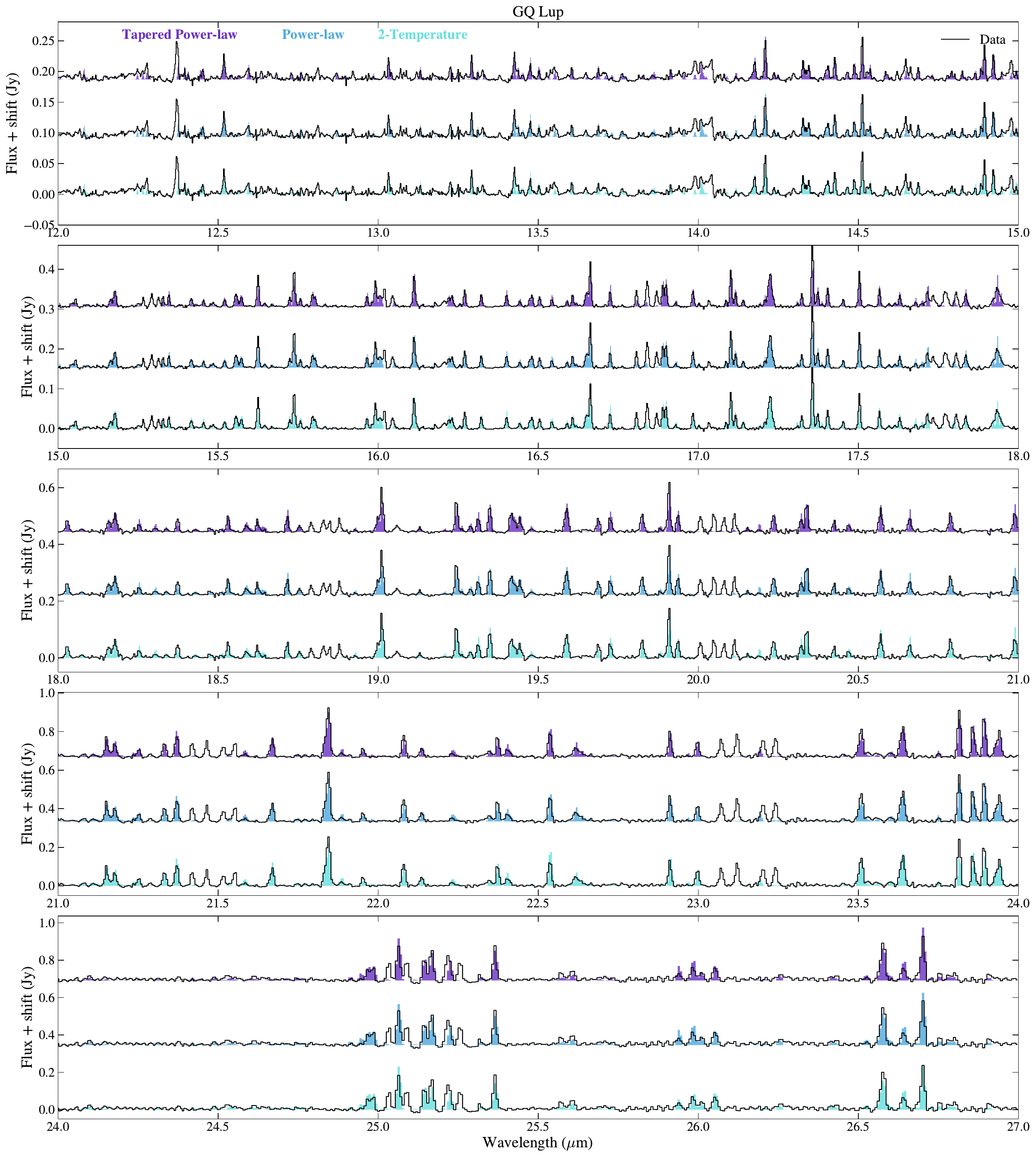}
    \caption{Same as Figure \ref{fig:FZTau_compare_models}, for GQ Lup.}
    \label{fig:GQLup_compare_models}
\end{figure*}

\begin{figure*}[ht!]
    \centering
    \includegraphics[width=0.98\textwidth]{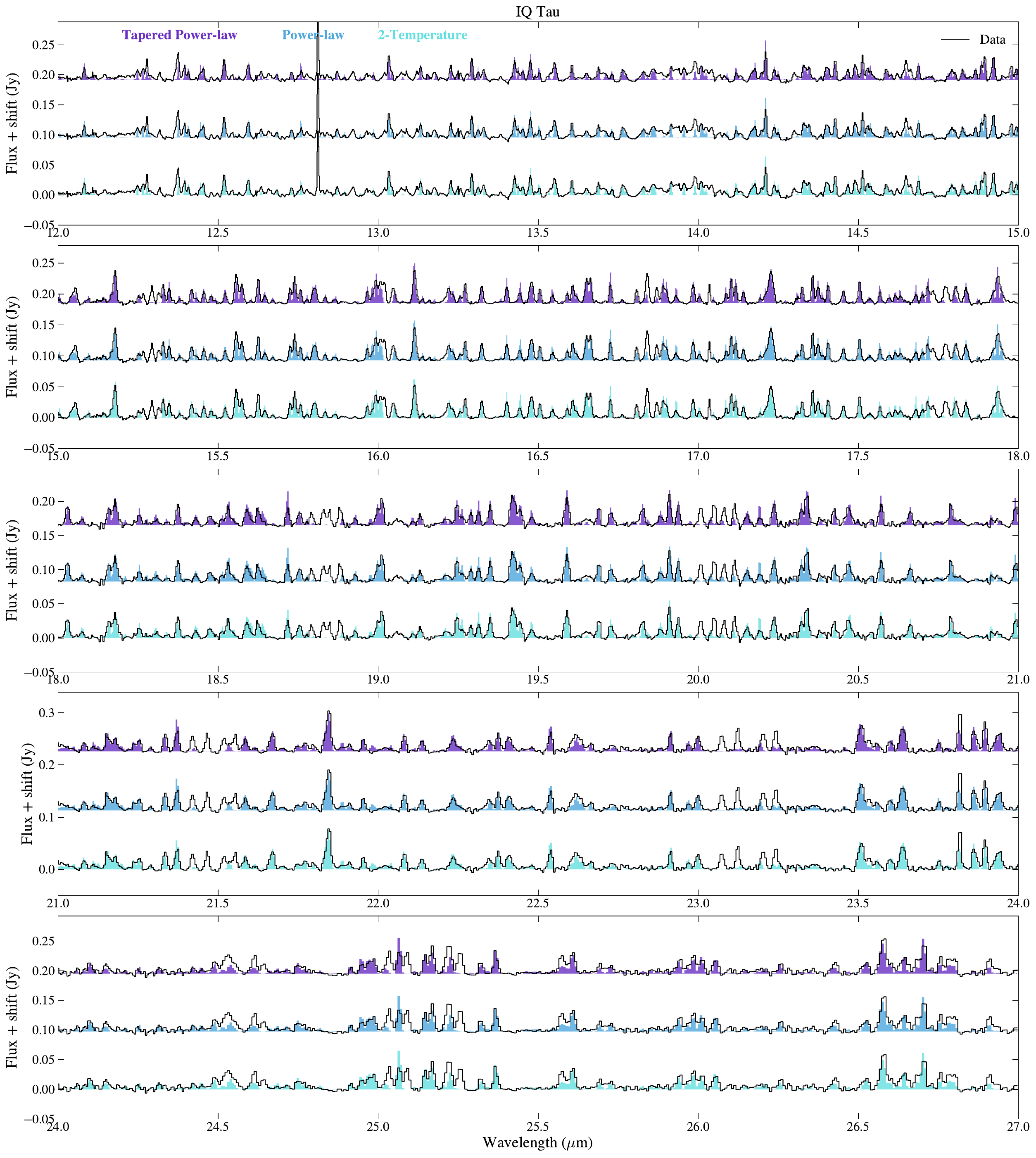}
    \caption{Same as Figure \ref{fig:FZTau_compare_models}, for IQ Tau.}
    \label{fig:IQTau_compare_models}
\end{figure*}

\begin{figure*}[ht!]
    \centering
    \includegraphics[width=0.98\textwidth]{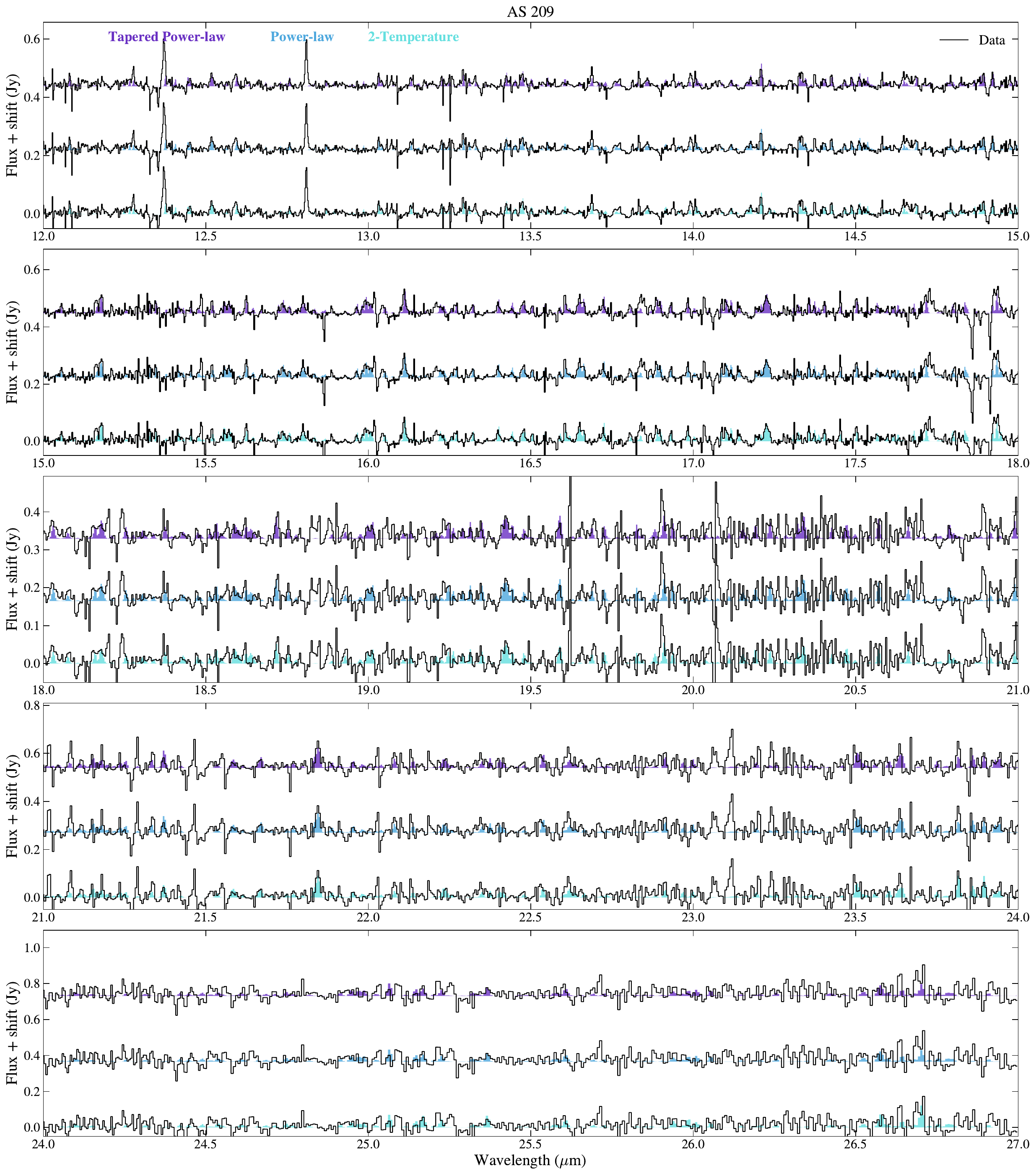}
    \caption{Same as Figure \ref{fig:FZTau_compare_models}, for AS 209.}
    \label{fig:AS209_compare_models}
\end{figure*}

\begin{figure*}[ht!]
    \centering
    \includegraphics[width=0.98\textwidth]{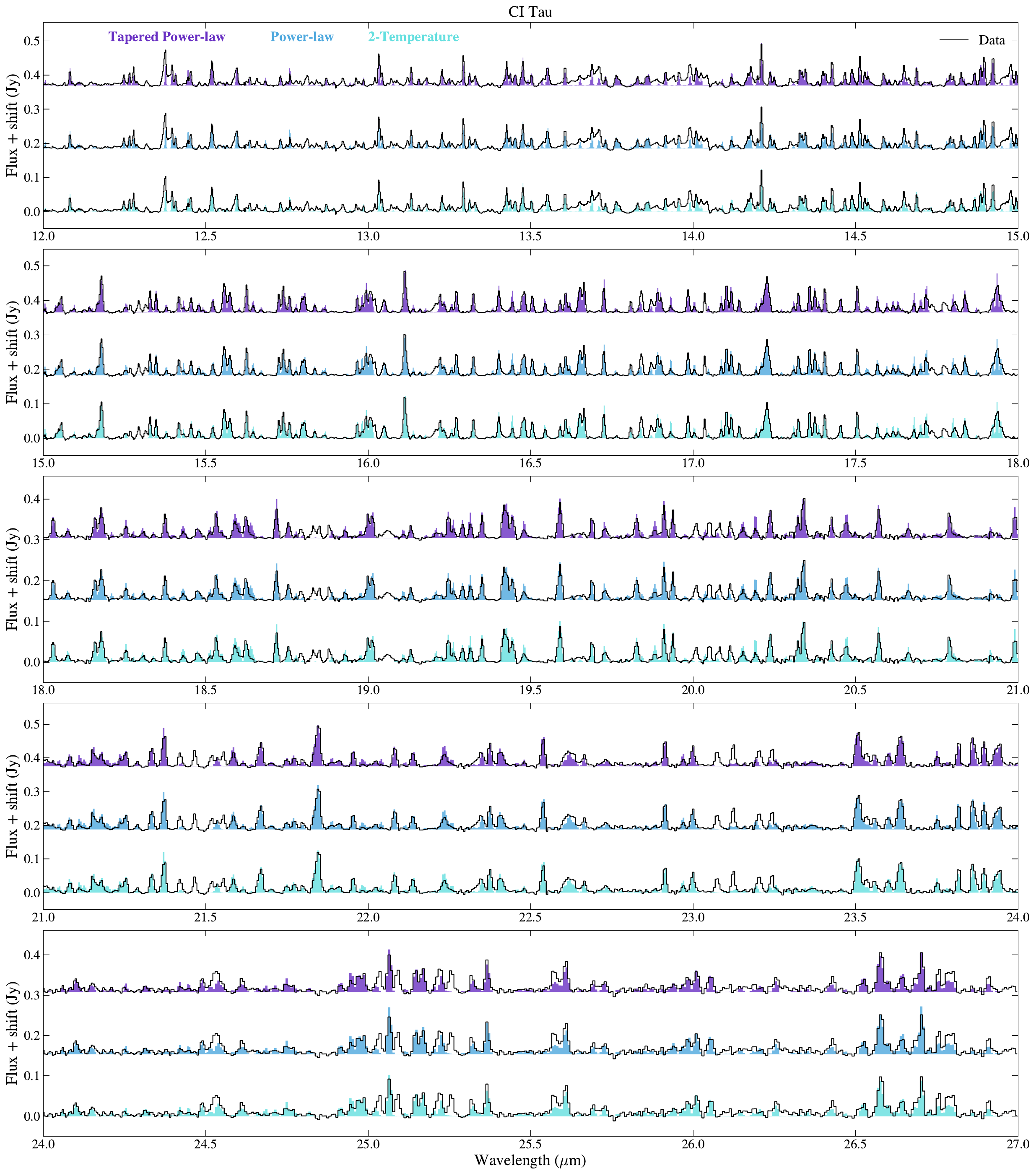}
    \caption{Same as Figure \ref{fig:FZTau_compare_models}, for CI Tau.}
    \label{fig:CITau_compare_models}
\end{figure*}

In Figures \ref{fig:GKTau_compare_models}-\ref{fig:CITau_compare_models} we present the comparison between two-temperature, power law, and tapered power law models for the five disks not shown in Section \ref{sec:mod_and_res}.

\section{Column Density and Emitting Area Correlations}
\label{apx:extra_corr}

\begin{figure*}[ht!]
    \centering
    \includegraphics[width=0.95\textwidth]{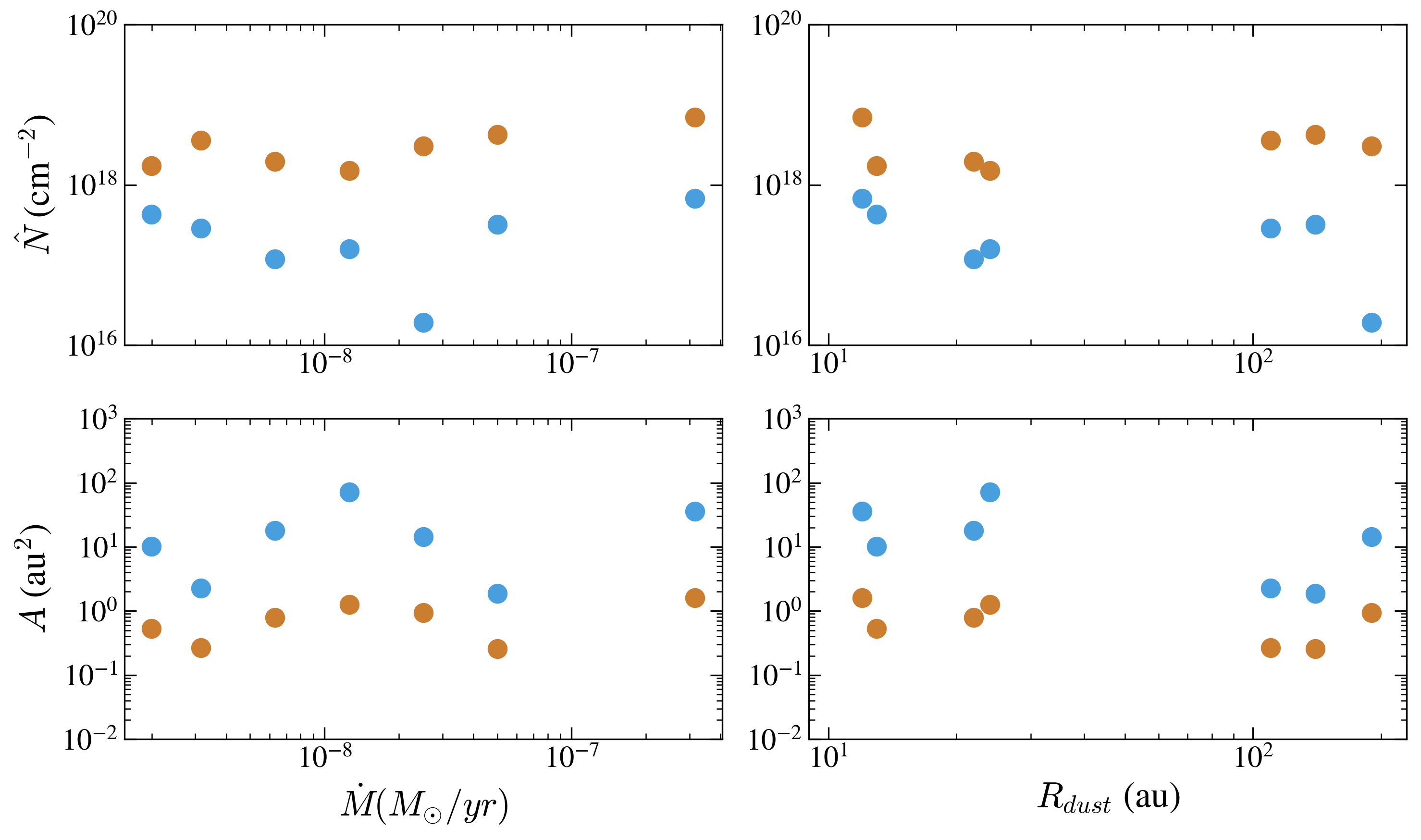}
    \caption{\textit{Top:} the average column density for cold (blue) and hot (orange) water vapor, extracted from the tapered column density profiles, as a function of mass accretion rate (left) and sub-mm dust disk radius (right). \textit{Bottom:} same as top panel, for the total emitting area of hot and cold water vapor.}
    \label{fig:tapered_coldhotNA}
\end{figure*}

 In Figure \ref{fig:tapered_coldhotNA} we show the average column density and emitting area for hot and cold water vapor plotted against $R_{dust}$ and $\dot{M}$. We note that neither parameter presents a statistically significant correlation with either the mass accretion rate or sub-mm dust disk radius, so the correlations reported for the observable mass appear to arise rather from the interplay of both parameters.

\section{Example Posterior Distributions}
\label{apx:posteriors}

\begin{figure*}[ht!]
    \centering
    \includegraphics[width=0.98\textwidth]{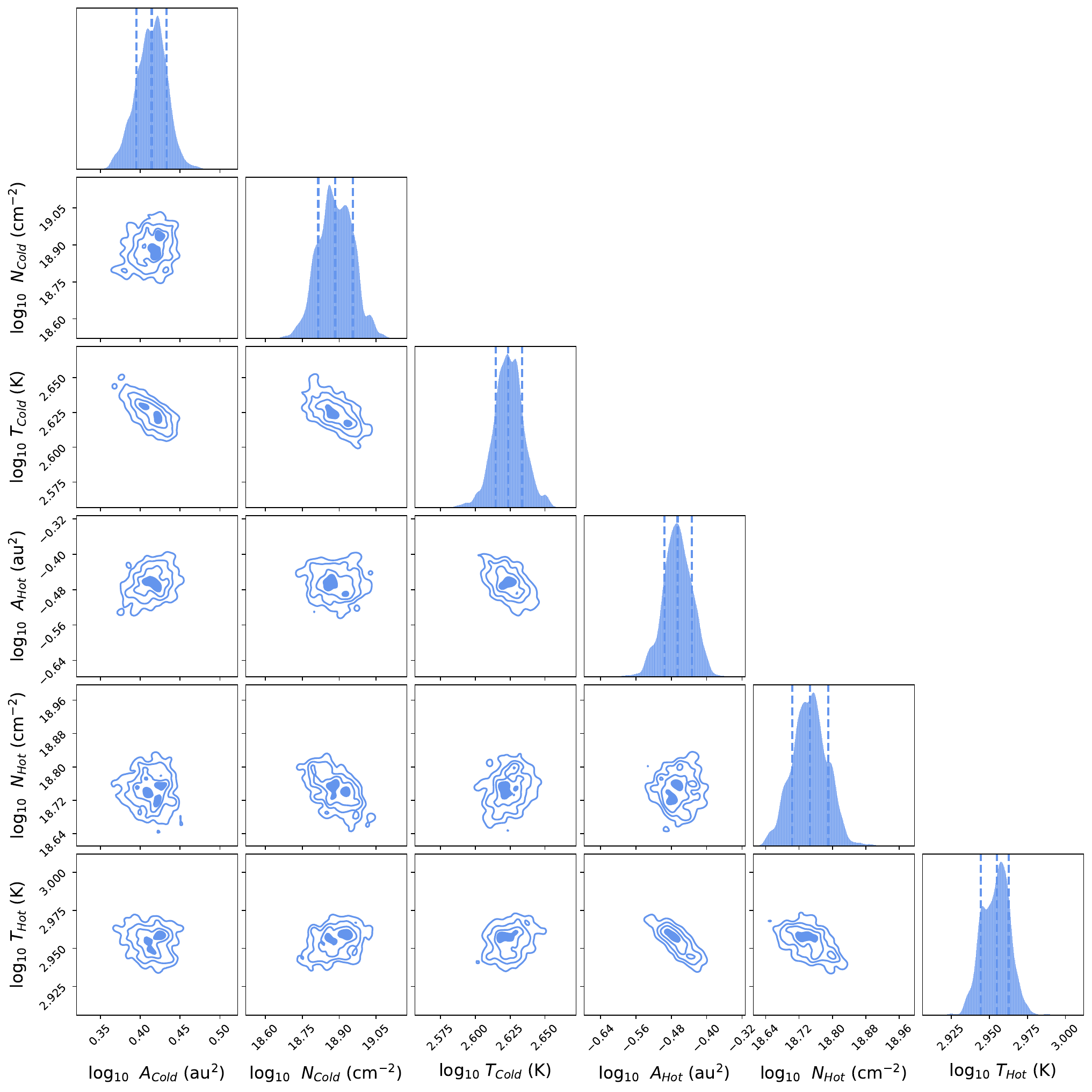}
    \caption{Corner plot for the two-temperature fit to FZ Tau. Dashed lines indicate the $1\sigma$ percentiles. The 2D histogram contours correspond to the 0.5, 1.0, 1.5, and 2.0 $\sigma$ confidence levels. The histograms have been smoothed by a Gaussian kernel of width 0.02$\%$ of the prior span.}
    \label{fig:FZTau_2comp_posterior}
\end{figure*}

\begin{figure*}[ht!]
    \centering
    \includegraphics[width=0.7\textwidth]{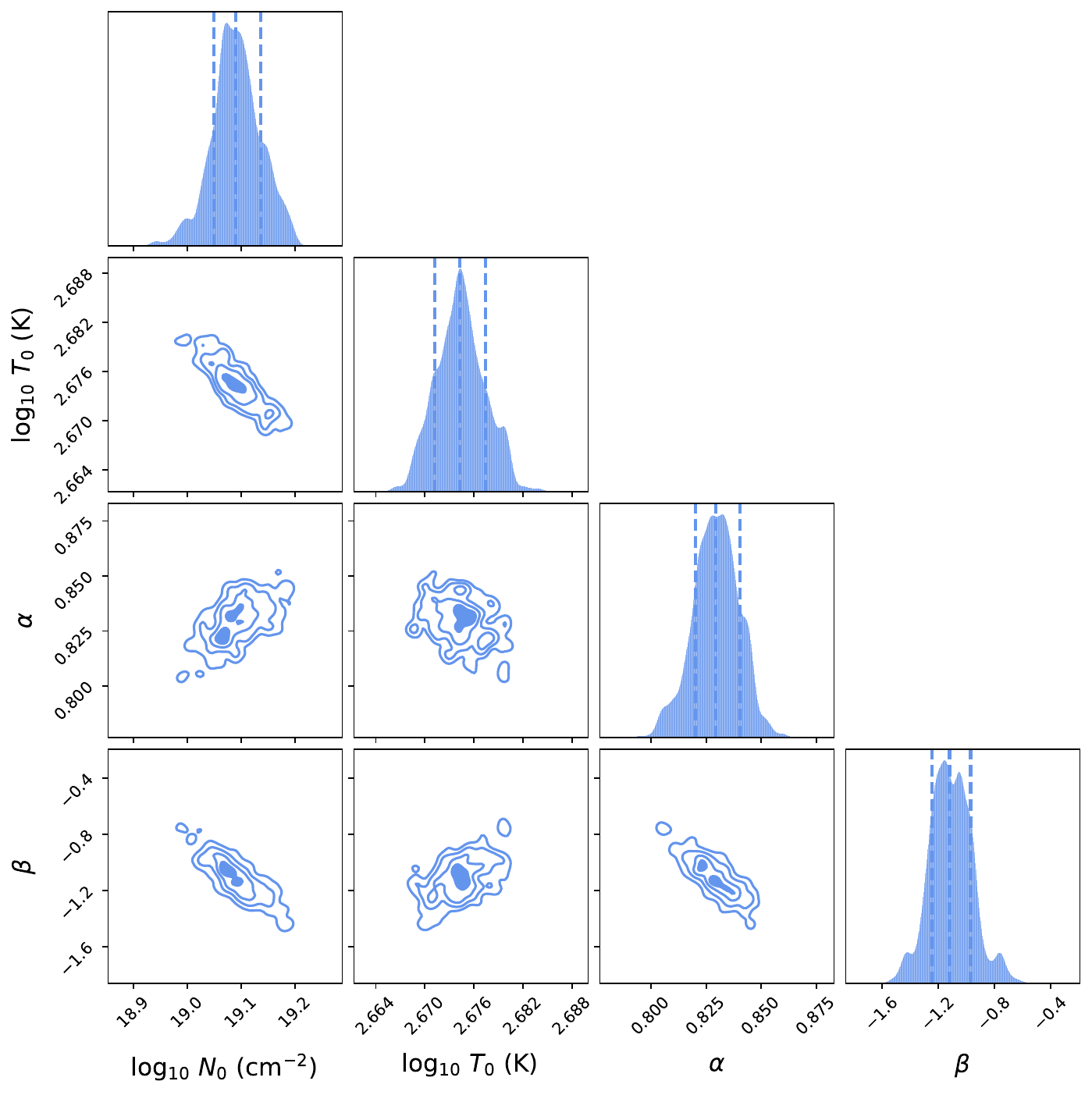}
    \caption{Same as Figure \ref{fig:FZTau_2comp_posterior}, for the power law column density fit.}
    \label{fig:FZTau_powerlaw_posterior}
\end{figure*}

\begin{figure*}[ht!]
    \centering
    \includegraphics[width=0.98\textwidth]{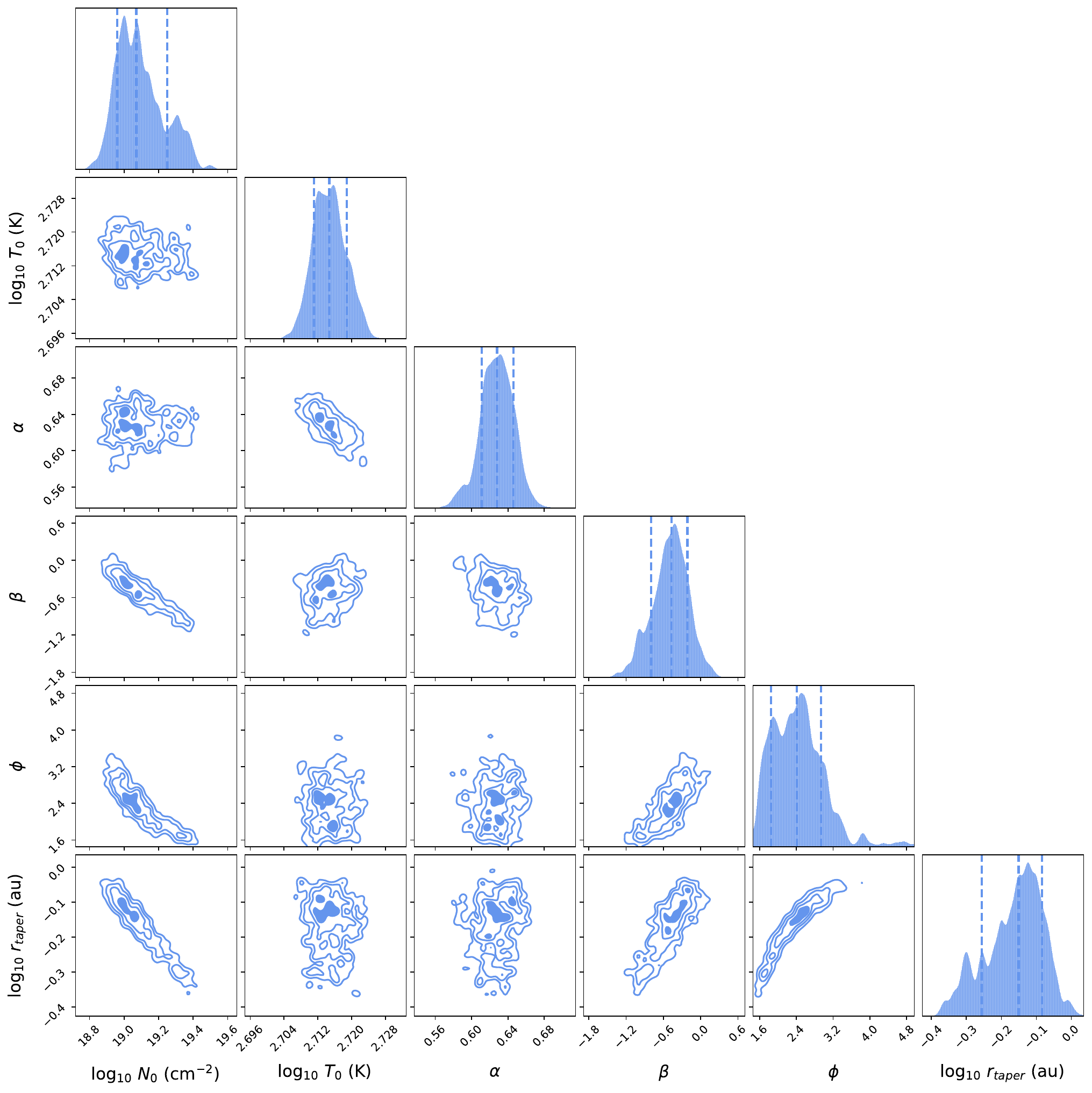}
    \caption{Same as Figure \ref{fig:FZTau_2comp_posterior}, for the exponentially tapered power law column density fit.}
    \label{fig:FZTau_taper_freer_posterior}
\end{figure*}

In Figures \ref{fig:FZTau_2comp_posterior}, \ref{fig:FZTau_powerlaw_posterior}, and \ref{fig:FZTau_taper_freer_posterior} we present corner plots with the posterior distributions obtained for each of the three fits (two-temperature, power law, and tapered power law) carried out for the disk of FZ Tau. 

For the two-component and power-law posterior distributions, we do not observe strong correlations. We note, however, that often the tapered power-law posterior distributions exhibit correlations between parameters, namely those of the tapered column density profiles. This is because multiple combinations of parameters result in almost identical profiles (as shown in Figure \ref{fig:tapered_profiles}, the column density profiles themselves are very well constrained). These correlations thus do not affect the results and conclusions presented in this paper, but only the interpretability of the some of the column density profile parameters.

\bibliography{sample631}{}
\bibliographystyle{aasjournal}

\end{document}